\documentclass[aps,prd,preprint,eqsecnum,tightenlines,floats,
showpacs,superscriptaddress]{revtex4}
\usepackage{amssymb} 
\usepackage{hyperref} 
\usepackage{graphicx}
\usepackage{subfigure}

\def\UF{University of Florida, Gainesville, FL 32611, USA}

\begin{document}

\preprint{LAUR 08-0789, UFIFT-HET-08-8}

\date{May 28, 2008}

\title{The Caustic Ring Model of the Milky Way Halo}

\author{L. D. Duffy}
\email{lduffy@lanl.gov}
\affiliation{Theoretical Division, Los Alamos National Laboratory, 
Los Alamos, NM 87545, USA}
\author{P. Sikivie}
\email{sikivie@phys.ufl.edu}
\affiliation{\UF}

\begin{abstract}

We present a proposal for the full phase space distribution of the 
Milky Way halo.  The model is axially and reflection symmetric and 
its time evolution is self-similar.  It describes the halo as a set 
of discrete dark matter flows with stated densities and velocity 
vectors everywhere.  We first discuss the general conditions under 
which the time evolution of a cold collisionless self-gravitating 
fluid is self-similar, and show that symmetry is not necessary for 
self-similarity.  When spherical symmetry is imposed, the model is 
the same as described by Fillmore and Goldreich, and by Bertschinger, 
twenty-three years ago.  The spherically symmetric model depends on 
one dimensionless parameter $\epsilon$ and two dimensionful parameters.  
We set $\epsilon$ = 0.3, a value consistent with the slope of the 
power spectrum of density perturbations on galactic scales.  The
dimensionful parameters are determined by the Galactic rotation 
velocity (220 km/s) at the position of the Sun and by the age of 
the Galaxy (13.7 Gyr).  The properties of the outer caustics are 
derived in the spherically symmetric model.  The structure of the 
inner halo depends on the angular momentum distribution of the dark 
matter particles.  We assume that distribution to be axial and 
reflection symmetric, and dominated by net overall rotation.  The 
inner caustics are rings whose radii are determined in terms of a 
single additional parameter $j_{\rm max}$.  We summarize the 
observational evidence in support of the model.  The evidence is 
consistent with $j_{\rm max}$ = 0.18 in Concordance Cosmology, 
equivalent to $j_{\rm max,old}$ = 0.26 in Einstein - de Sitter 
cosmology.  We give formulas to estimate the flow densities and 
velocity vectors anywhere in the Milky Way halo.   The properties 
of the first forty flows at the location of the Earth are listed.

\end{abstract}

\maketitle

\section{Introduction}

It has been established from a variety of observational inputs 
that approximately 23\% of the energy density of the universe 
is in the form of ``cold dark matter" (CDM)\cite{WMAP}.  The CDM 
particles must be non-baryonic, collisionless, and cold.  ``Cold" 
means that their primordial velocity dispersion is small enough 
that it can be set equal to zero for all practical purposes when 
discussing the formation of large scale structure.  ``Collisionless" 
means that the particles have negligible interactions other than 
gravity.  A central problem in dark matter studies is the question 
how CDM is distributed in the halos of galaxies, and in particular 
in the halo of our Milky Way galaxy.  Indeed, knowledge of this 
distribution is essential for understanding galactic dynamics 
and for predicting signals in direct and indirect searches for 
dark matter on Earth.

Galactic halos are collisionless fluids and must therefore be 
described in 6-dimensional phase space.  A full description 
gives the phase space distribution $f(\vec{r}, \vec{v}; t)$
of the dark matter particles in the halo, i.e. their velocity 
($\vec{v}$) distribution at every position $\vec{r}$.  $t$ is 
time.  An important simplification occurs in the case of halos 
built of {\it cold} collisionless dark matter because CDM 
particles lie in phase space on a thin 3-dimensional hypersurface.  
This fact implies that the velocity distribution is everywhere 
discrete \cite{Ips} and that there are surfaces in physical 
space, called caustics, where the density of dark matter is 
very large.  See Fig. 1 for an illustration. It is argued in 
ref. \cite{rob} that discrete flows and caustics in galactic 
halos are a generic and robust prediction of cold dark matter 
cosmology, even after all possible complications and reasons 
for skepticism have been considered.  The reader may wish to 
consult ref.~\cite{rob} for background information and a list 
of references.  Discrete flows and caustics are seen in 
N-body simulations of large scale structure formation when 
special techniques are used to reach adequate resolution in 
the relevant regions of phase-space \cite{simca,Stiff,Vogel}.

The caustic ring model is a proposal for the full phase space
distribution of the Milky Way halo.  The model is axially symmetric, 
reflection symmetric, and self-similar in its time evolution.  The 
model is an elaboration of the spherically symmetric self-similar 
model originally proposed by Fillmore and Goldreich \cite{FG}, and 
by Bertschinger \cite{B}.  Our main purpose in presenting the caustic 
ring model is to enable its further comparison with observation.  The 
observational evidence which we claim in support of the model thus 
far is listed in Section IV.

Because the original self-similar halo model \cite{FG,B} is 
spherically symmetric, one may be tempted to suppose that 
self-similarity and spherical symmetry are linked somehow, 
that the latter is required by the former.  This is not so, 
however.  In Section II, we discuss the general conditions 
under which the time evolution of a cold collisionless 
self-gravitating fluid is self-similar.  We show that 
self-similarity does not require symmetry.  We consider 
three cases: no symmetry, spherical symmetry, and axial 
symmetry.  In each instance, we derive the equations of 
motion for the reduced variables, and the initial conditions 
required by self-similarity.

The spherically symmetric infall model is inadequate to describe 
the inner parts of a cold dark matter halo.  Indeed, in a spherically
symmetric halo, CDM particles necessarily have vanishing angular
momentum.  Each particle moves on a radial orbit and passes through 
the center at each throughfall.  The center is then a caustic point 
where the density $d(r)$ diverges approximately as ${1 \over r^2}$, 
where $r$ is galactocentric distance.  In realistic halos, the dark 
matter particles carry angular momentum and therefore miss the center.  
There are still inner caustics in that case \cite{inner} but they
are spread on surfaces with cusps.  The catastrophe structure of the 
inner caustics \cite {inner} depends on the angular momentum distribution 
of the infalling CDM particles.  If the angular momentum distribution 
is dominated by net overall rotation, the inner caustics are rings 
(closed tubes) whose cross-section is a section of the elliptic 
umbilic ($D_{-4}$) catastrophe \cite{crdm,sing}.  These objects are 
usually referred to as ``caustic rings of dark matter".  Figs.~\ref{flow}
and \ref{dimen} describe the cross-section of a caustic ring of dark 
matter and indicate what is meant by the ring radius $a$ and by its 
transverse dimensions $p$ and $q$.  In the opposite case, where the 
velocity distribution is irrotational ($\vec{\nabla} \times \vec{v} = 0$), 
the inner caustic has a tent-like structure described in detail in 
ref. \cite{inner}.  Because there is evidence for caustic rings of 
dark matter (as opposed to tent-like caustics), in the Milky Way
\cite{crdm,milk} and in other isolated spiral galaxies \cite{Kinn}, 
we assume that the angular momentum distribution is dominated 
by net overall rotation.

The spherically symmetric self-similar infall model \cite{FG,B} 
was generalized in ref. \cite{STW} to include angular momentum 
of the infalling particles.  The main purpose of that paper 
was to estimate the densities and speeds of the dark matter 
flows at a distance of 8.5 kpc from the Galactic center.  It 
was shown in ref. \cite{STW} that the angular momentum distribution 
can be chosen so as to preserve self-similarity.  For the sake of 
convenience, the model of ref. \cite{STW} was artificially made 
spherically symmetric by averaging an actual halo with angular
momentum over all orientations.  No such averaged halo can be 
physically realized in cold dark matter cosmology.  However, 
the model of ref. \cite{STW} is adequate for its purpose of 
estimating the speeds and average densities of the dark matter 
flows at 8.5 kpc from the Galactic center.  The  model of 
ref. \cite{STW} has inner caustics but they are spread over 
spheres.  The model of ref. \cite{STW} is an important precursor 
to the model presented here.  The latter is essentially the former 
without the averaging over all orientations. 

It was assumed in ref. \cite{STW} that the universe is 
Einstein - de Sitter, i.e. spatially flat and matter dominated 
($\Omega = \Omega_{\rm m} = 1$).  This is a convenient assumption 
when building a self-similar model because self-similarity 
requires the absence of a special time.  The Einstein - de Sitter 
cosmology doesn't have a special time and is the only cosmology in 
which structure formation can be strictly self-similar.  Subsequently,
the accelerated expansion of the universe was discovered \cite{accex} 
and the Concordance Cosmology ($\Omega_{\rm m}$ = 0.27 and 
$\Omega_\Lambda$ = 0.73) established \cite{WMAP}.  The 
Concordance Cosmology is inconsistent with exact self-similarity 
since it has a special time, namely the time a few Gigayears ago 
when the universe passed from matter domination to dark energy 
domination.  However, the Milky Way halo was assembled almost 
entirely before the onset of accelerated expansion, when the 
universe was well described by $\Omega = \Omega_{\rm m}$ = 1 
cosmology.  The breaking of self-similarity implied by the 
Concordance Cosmology has only a small effect on the formation 
of our Galaxy and will be ignored here.  

Nonetheless, the discovery of the accelerated expansion of the 
universe and the advent of Concordance Cosmolgy did affect the 
self-similar infall model in an important way because they led 
to a precise determination of the age of the universe.  In 
ref.~\cite{STW} it was assumed, consistently with Einstein 
- de Sitter cosmology, that the present age is related to 
the present Hubble expansion rate $H_0$ by  
$t_0 = {2 \over 3 H_0} = {1 \over h}$ 6.52 Gyr, where $h$ 
is defined as usual by $H_0 = h$ 100 km/(s Mpc).  Since the 
universe is now known to be 13.7 Gyr old \cite{accex,WMAP}, 
the results of ref.~\cite{STW} remain valid provided only 
that $h$ is replaced everywhere in that paper by 
${6.52~{\rm Gyr} \over t_0}$ = 0.476.

The outline of our paper is as follows.  In Section II, we 
discuss the general conditions under which the time evolution 
of a cold collisionless self-gravitating fluid is self-similar.
We show that spatial symmetry is not necessary for self-similarity.
We discuss three cases - no symmetry, spherical symmetry and 
axial symmetry - and obtain the equations of motion for the 
reduced variables in each case.  In Section III, we present 
the model.  We discuss its underlying assumptions, and describe 
its outer and inner caustics.  We provide interpolating formulae
to estimate the flow velocities and densities between the inner 
and outer caustics.  We list the estimates for the inner and 
outer caustic properties which follow from the application of 
the method of adiabatic invariants to the model.  In Section IV, 
we list the evidence in support of the model.  In Section V, we 
list the densities and velocities of the first forty flows on 
Earth.  Section VI provides a summary.

\section{Self-similar cold dark matter halos}

The purpose of this section is to obtain the conditions under 
which the time evolution of a cold self-gravitating collisionless
fluid is self-similar.  We show that symmetry is not necessary for
self-similarity.  First, we discuss {\it self-similar} collisionless 
fluids.  Next, we discuss {\it cold} collisionless fluids.  Thirdly, 
we discuss {\it cold and self-similar} collisionless fluids. In each 
instance we obtain the relevant equations, without assumptions of 
symmetry.  Finally, we specialize to the cases of spherical and 
axial symmetry.

\subsection{General conditions for self-similarity}

Let $f(\vec{r}, \vec{v}; t)$ be the phase space density of 
a collisionless self-gravitating fluid at time $t$.  It 
satisfies the collisionless Boltzmann equation
\begin{equation}
{df \over dt} = {\partial f \over \partial t} +
\vec{v}\cdot{\partial f \over \partial \vec{r}} -
\vec{\nabla}\Phi \cdot {\partial f \over \partial \vec{v}} = 0~~~~\ ,
\label{Bol}
\end{equation}
where
\begin{equation}
\Phi(\vec{r}, t) = - G m \int d^3 r^\prime \int d^3 v~
{f(\vec{r}~^\prime,\vec{v}; t) \over |\vec{r} - \vec{r}~^\prime|}
\label{pot}
\end{equation}
is the gravitational potential. $m$ is the particle mass.

Let us first discuss simple rescalings of the phase space 
distribution and evolution rate of such a system.  Given 
that $f(\vec{r}, \vec{v}; t)$ is a solution of Eqs.~(\ref{Bol})
and (\ref{pot}), consider
\begin{equation}
f^\prime(\vec{r}, \vec{v}; t) = A 
f({\vec{r} \over R}, {\vec{v} \over V}; {t \over T})~~~~\ ,
\label{fprim}
\end{equation}
where $A$, $R$, $V$ and $T$ are constants.  It is straightforward 
to show that $f^\prime(\vec{r}, \vec{v}; t)$ is also a solution 
provided
\begin{equation}
V = {R \over T}~~~~~{\rm and}~~~~~A = {T \over R^3}~~~~~\ .
\label{cond}
\end{equation}
Therefore any solution can always be rescaled in two ways: it 
can be made older (and hence more slowly evolving) and it can 
be made larger.  The age $t$ sets the scale for the density 
$\rho$ since $G \rho \sim {1 \over t^2}$.  The size $R$ then 
sets the scale for the velocities $v \sim {R \over t}$, and 
for the total mass $M \sim \rho R^3 \sim {R^3 \over t^2}$.  
Next let us consider self-similar evolution of the fluid, 
which is an invariance under time-dependent rescalings.

The evolution is called self-similar if, in all aspects, the 
fluid remains identical to itself except for an overall rescaling 
of its phase space density by a factor $A(t)$, of its size in the 
spatial directions by a factor $R(t)$ and of its size in the velocity
directions by a factor $V(t)$.  Thus, the phase space density of 
a fluid with self-similar evolution satisfies the ansatz
\begin{equation}
f(\vec{r}, \vec{v}; t) = A(t)
F\left({\vec{r} \over R(t)}, {\vec{v} \over V(t)}\right)
\label{ans}
\end{equation}
where $F(\vec{\chi}, \vec{\nu})$ is a rescaled time-independent
phase space density.  The gravitational potential is then
\begin{equation}
\Phi(\vec{r}, t) = A(t) R(t)^2 V(t)^3~
\Psi\left({\vec{r} \over R(t)}\right)
\label{Phi}
\end{equation}
with
\begin{equation}
\Psi(\vec{\chi}) = - G m \int d^3 \chi^\prime \int d^3 \nu~
{F(\vec{\chi}^\prime, \vec{\nu}) \over 
|\vec{\chi} - \vec{\chi}^\prime|}~~~\ .
\label{Psi}
\end{equation}
Substituting Eqs.~(\ref{ans}) and (\ref{Phi}) into 
Eq.~(\ref{Bol}), one finds that the self-similarity 
ansatz can be satisfied only if 
\begin{equation}
R(t) \propto t^\beta~~~,~~~
V(t) = {R(t) \over t} ~~~{\rm and}~~~
A(t) = {t \over R(t)^3}~~~\ .  
\label{pow}
\end{equation}
The rescaled phase space distribution must satisfy
\begin{equation}
(1 - 3 \beta) F(\vec{\chi}, \vec{\nu}) +
(\vec{\nu} - \beta \vec{\chi}) \cdot {\partial F \over \partial\vec{\chi}}
+ \left((1 - \beta)\vec{\nu} - 
{\partial \Psi \over \partial\vec{\chi}}\right) 
\cdot {\partial F \over \partial\vec{\nu}} = 0~~~~\ .
\label{Feq}
\end{equation}
Thus, the evolution of a self-similar self-gravitating collisionless
fluid is obtained by choosing $\beta$ and solving simultaneously
Eqs.~(\ref{Feq}) and (\ref{Psi}), with appropriate boundary 
conditions.  Next, we consider a collisionless fluid which 
is self-gravitating and cold, but not necessarily self-similar.

\subsection{Phase space distribution of cold dark matter}

Cold collisionless dark matter (CDM) particles, such as axions or 
WIMPs, lie on a thin 3-dim. hypersurface in phase space.  Indeed, at 
a sufficiently early time $t_{\rm in}$, all CDM particles at a given 
location $\vec{r}$ have the same velocity $\vec{v}_{\rm in}(\vec{r})$, 
up to a small species-dependent primordial velocity dispersion $\delta v$.  
We call the 3-dim. hypersurface the ``phase space sheet".  The thickness
of the phase space sheet is the primordial velocity dispersion $\delta v$.  
For CDM $\delta v$ is small, i.e. the sheet is thin \cite{sing}.  The
number of particles is huge (approx. $10^{84}$ axions and/or $10^{68}$
WIMPs per galactic halo), so that the sheet is continuous.  As time 
goes on, the sheet folds in phase space.

To describe the evolution of the phase space sheet 
we label the particles by a continuous parameter 
$\vec{\alpha} =(\alpha_1, \alpha_2, \alpha_3)$.  
The position of particle labeled $\vec{\alpha}$ at 
time $t$ is $\vec{x}(\vec{\alpha}, t)$.  The particle 
has velocity $\vec{v}(\vec{\alpha}, t) = 
{\partial \vec{x} \over \partial t}(\vec{\alpha}, t)$.  
The phase space sheet is the set of points
$(\vec{x}(\vec{\alpha},t), \vec{v}(\vec{\alpha}, t))$ 
for all $~\vec{\alpha}$.  It has a time-independent mass
density ${d M \over d \alpha^3}(\vec{\alpha})$ in parameter
space.  At any point $\vec{r}$ in physical space, there is a 
discrete set of flows.  The number $N(\vec{r}, t)$ of flows 
at position $\vec{r}$ and time $t$ is the number of solutions 
$\vec{\alpha}_j(\vec{r}, t)$, $j = 1, 2 ... N$, of the equation 
$\vec{r} = \vec{x}(\vec{\alpha}, t)$.  The $j^{\rm th}$ flow has 
velocity
\begin{equation}
\vec{v}_j(\vec{r}, t) = 
{\partial \vec{x} \over \partial t}(\vec{\alpha}_j(\vec{r}, t), t)
\label{vel}
\end{equation}
and density
\begin{equation}
d_j(\vec{r}, t) = 
{d M \over d \alpha^3}(\vec{\alpha}_j(\vec{r}, t))
{1 \over |D(\vec{\alpha}_j(\vec{r}, t), t)|}
\label{den}
\end{equation}
where
\begin{equation}
D(\vec{\alpha}, t) = 
det \left({\partial \vec{r} (\vec{\alpha}, t) 
\over \partial\vec{\alpha}} \right)~~~~\ .
\label{Jac}
\end{equation}
Caustics are located where $D(\vec{\alpha}_j(\vec{r}, t), t) = 0$ for 
some $j$.  A possible parametrization is $\vec{\alpha} = \vec{r}_{\rm in}$
where $\vec{r}_{\rm in}$ is particle position at an arbitrarily chosen but
sufficiently early initial time $t_{\rm in}$.  Other parametrizations may 
be more convenient, however.  The flow velocities $\vec{v}_j(\vec{r}, t)$ 
and densities $d_j(\vec{r}, t)$ are parametrization independent.  

Since the particles are collisionless, 
\begin{equation}
{\partial^2 \vec{x}(\vec{\alpha}, t) \over \partial t^2} = 
- \vec{\nabla} [\Phi(\vec{r}, t) +
\Phi_{\rm b}(\vec{r}, t)]|_{\vec{r} = \vec{x}(\vec{\alpha}, t)}
\label{New}
\end{equation}
where
\begin{equation}
\Phi(\vec{r}, t) = 
- G \int d^3 r^\prime \sum_{j = 1}^{N(\vec{r}^\prime, t)}
{d_j(\vec{r}^\prime, t) \over |\vec{r} - \vec{r}^\prime|}
= - G \int d^3 \alpha {d M \over d \alpha^3}(\vec{\alpha})
{1 \over |\vec{r} - \vec{x}(\vec{\alpha}, t)|}
\label{cPhi}
\end{equation}
is the gravitational potential due to dark matter and 
$\Phi_{\rm b}$ is the contribution of ordinary matter.  
The CDM phase space distribution is determined by 
specifying ${d M \over d \alpha^3}(\vec{\alpha})$ 
and the positions $\vec{r}_{\rm in}(\vec{\alpha})$ 
and velocities $\vec{v}_{\rm in}(\vec{\alpha})$ of 
the particles at an initial time $t_{\rm in}$, and 
solving Eqs.~(\ref{New}) and (\ref{cPhi}) simultaneously.  
$\Phi_b(\vec{r}, t)$ has to be supplied by hand or derived 
in some way.

We wish to apply the above to the construction of 
galactic halos.  Galactic halos grow by accreting more
and more of the surrounding dark matter.  A convenient 
choice for the parameters $\vec{\alpha}$ is as follows.
Define a sphere of radius $R(t)$ which grows with the 
halo.  $R(t)$ is chosen such that each particle traverses
the sphere only once, in the inward direction.  The particles
are labeled by $\vec{\alpha} = (\theta, \varphi, \tau)$ where
$(\theta, \varphi)$ are the spherical coordinates of the point
where the particle crosses the sphere and $\tau$ is the time 
when it does so.  For given $R(t)$, the model halo is specified 
by giving the velocities $\vec{v}_{\rm in}(\theta, \varphi, \tau)$ 
of the particles when they cross the sphere and their infall 
rate per unit solid angle:
\begin{equation}
{d M \over d\Omega d\tau}(\theta, \varphi, \tau) = 
{1 \over \sin \theta} {d M \over d\theta d\varphi d\tau}
(\theta, \varphi, \tau)~~~~~\ .
\label{inr}
\end{equation}
Note that the number of degrees of freedom in 
$\vec{v}_{\rm in}(\theta, \varphi, \tau)$ is slightly
reduced by the freedom to choose $R(t)$.  For example, 
we may choose $R(t)$ such that the particle which is 
at the intersection of the sphere with the $x$-axis
is at first turnaround, for all $t$.  In that case, 
$\vec{v}_{\rm in}({\pi \over 2}, 0, \tau)\cdot\hat{x} 
= 0$ for all $\tau$.

The gravitational potential due to dark matter may be 
separated into two parts: 
$\Phi = \Phi_{\rm in} + \Phi_{\rm ex}$, with 
\begin{equation}
\Phi_{\rm in}(\vec{r}, t) = - G 
\int d\Omega \int_0^t d\tau {d M \over d\Omega d\tau} 
{1 \over |\vec{r} - \vec{x}(\theta, \varphi, \tau; t)|}
\label{in}
\end{equation}
the contribution of particles that have already fallen 
through the sphere, and
\begin{equation}
\Phi_{\rm ex}(\vec{r}, t) = - G  
\int d\Omega \int_t^\infty d\tau {d M \over d\Omega d\tau}
{1 \over |\vec{r} - \vec{x}(\theta, \varphi, \tau; t)|}
\label{ex}
\end{equation}
the contribution of particles that have not yet done so. 
In the spherically symmetric case $\Phi_{\rm ex}$ can be 
ignored since it is $\vec{r}$-independent inside the sphere.  
In the non-spherically symmetric case, to obtain $\Phi_{\rm ex}$, 
the equations of motion have to be - in principle - integrated 
backward in time to determine the positions of the particles 
outside the sphere. This procedure would then include all the 
particles in the universe.  In practice, the contribution of 
particles outside the sphere may be given by hand, or ignored.  

\subsection{Self-similar evolution of cold dark matter}

In this subsection, we set $\Phi_b = 0$ because the presence 
of ordinary matter generally spoils exact self-similarity by 
introducing special time and special length scales.  However 
later, when discussing our Milky Way halo model, we will add 
the baryon contribution to the gravitational potential by hand, 
maintaining self-similarity, as described in Section III.

We saw in subsection II.A that a self-similar halo grows 
in physical space as a power law: $R(t) \propto t^\beta$.
Hence we choose the sphere introduced in subsection II.B to 
have radius
\begin{equation}
R(t) = R_0 \left({t \over t_0}\right)^\beta
\label{rad}
\end{equation}
where $t_0$ is an arbitrary reference time.  Under the 
self-similarity assumption, the particle that falls in 
at time $\tau$ and location $(\theta, \varphi)$ moves in 
exactly the same way as the particle that falls in at 
$(\theta, \varphi)$ at any other time $\tau^\prime$, except 
that its trajectory is rescaled in physical space by 
${R(\tau) \over R(\tau^\prime)}$ and in time by 
${\tau \over \tau^\prime}$.  Hence there exists a 
function $\vec{\lambda}(\theta, \varphi; \xi)$ such that 
\begin{equation}
\vec{x}(\theta, \varphi, \tau; t) = R(\tau)
\vec{\lambda}(\theta, \varphi; \xi = {t \over \tau})~~~~\ .
\label{selx}
\end{equation}
Also the infall rate ${dM \over d\Omega d\tau}(\theta, \varphi, \tau)$
must have a specific time dependence.  The total dark matter mass within
the sphere of radius $R(t)$ is 
\begin{eqnarray}
M(t) &=& \int d\Omega \int_0^t d\tau
{d M \over d\Omega d\tau}(\theta, \varphi, \tau)
= m \int_{r < R(t)} d^3 r \int d^3 v f(\vec{r}, \vec{v}; t)\nonumber\\
&=& m A(t) {R^6(t) \over t^3} \int_{|\vec{\chi}| < 1} d^3 \chi
\int d^3 \nu F(\vec{\chi}, \vec{\nu}) \propto t^{3\beta-2}~~~\ ,
\label{prop}
\end{eqnarray}
where we used Eqs.~(\ref{pow}).  Hence we need
\begin{equation}
{d M \over d\Omega d\tau}(\theta, \varphi, \tau) 
= \left({\tau \over t_0}\right)^{3(\beta-1)}
{d M \over d\Omega d\tau}\Bigg|_0(\theta, \varphi)~~~\ .
\label{infr}
\end{equation}
Finally, the initial velocities must be of the form
\begin{equation}
\vec{v}_{\rm in}(\theta, \varphi, \tau) = {R(\tau) \over \tau}
\vec{\nu}_{\rm in}(\theta, \varphi)
\label{vin}
\end{equation}
to be consistent with Eq. (\ref{selx}).

Substituting Eqs. (\ref{selx}) and (\ref{infr}) into Eq. (\ref{cPhi}), 
we obtain 
\begin{equation}
\Phi(\vec{r}, t) = \left({R(t) \over t}\right)^2
\Psi\left({\vec{r} \over R(t)}\right)
\label{nPhi}
\end{equation}
with
\begin{equation}
\Psi(\vec{\chi}) = - G \left({t_0 \over R_0}\right)^3
\int_0^\infty {d \xi \over \xi^{3\beta-1}} \int d\Omega
{d M \over d\Omega d\tau}\Bigg|_0(\theta, \varphi)
{1 \over |\vec{\chi} - 
{1 \over \xi^\beta}\vec{\lambda}(\theta,\varphi; \xi)|}~~~\ .
\label{nPsi}
\end{equation}
With $\Phi_{\rm b} = 0$, the equations of motion (\ref{New}) become 
\begin{equation}
{\partial^2 \vec{\lambda} \over \partial \xi^2} = 
- \xi^{\beta-2} {\partial \Psi \over \partial \vec{\chi}}
\left(\vec{\chi} = 
{\vec{\lambda}(\theta, \varphi; \xi) \over \xi^\beta}\right)~~~\ .
\label{eom}
\end{equation}
The initial conditions are 
\begin{eqnarray}
\vec{\lambda}(\theta, \varphi; 1) &=& \hat{n} (\theta, \varphi)\nonumber\\
{\partial \vec{\lambda} \over \partial \xi}(\theta, \varphi; 1) &=&
\vec{\nu}_{\rm in}(\theta, \varphi)~~~\ ,
\label{inc}
\end{eqnarray}
where $\hat{n}(\theta, \varphi)$ is the unit vector in the direction 
$(\theta, \varphi)$.  The halo is determined in terms of $\beta$, 
$\vec{\nu}_{\rm in} (\theta, \varphi)$ and 
${d M \over d\Omega d\tau}|_0 (\theta, \varphi)$ by solving 
Eqs. (\ref{nPsi}) and (\ref{eom}) with initial conditions (\ref{inc}).
Again, the freedom to choose $R$ removes one degree of freedom
from $\vec{\nu}_{\rm in}(\theta, \varphi)$.  For example, $R$ may be 
chosen such that $\vec{\nu}_{\rm in}({\pi \over 2}, 0)\cdot\hat{x} = 0$.

\subsection{The spherically symmetric case}

For a halo of cold collisionless particles to be spherically 
symmetric, the particle orbits must be radial, i.e. each particle 
has zero angular momentum.  Thus
\begin{equation}
\vec{\nu}_{\rm in} (\theta, \varphi) = \nu_{\rm in}
\hat{n}(\theta, \varphi)~~~~\ .
\label{sphn}
\end{equation}
Furthermore $\nu_{\rm in}$ and ${d M \over d\Omega d\tau}|_0$ must 
be independent of $(\theta, \varphi)$.  We choose $R(t)$ to be the 
turnaround radius at time $t$, so that $\nu_{\rm in} = 0$.  We 
have then
\begin{equation}
\Psi(\chi) = - G \left({t_0 \over R_0}\right)^3 {dM \over dt}(t_0)
\int_0^\infty {d \xi \over \xi^{3\beta-1}} 
{1 \over |\chi - {1 \over \xi^\beta} \lambda(\xi)|}~~~\ .
\label{sphp}
\end{equation}
Because of spherical symmetry, 
${\partial \Psi \over \partial \chi}(\chi)$ gets 
contributions only from $\xi$ such that 
${1 \over \xi^\beta} \lambda(\xi) < \chi$.  In particular, 
for $\chi < 1$ there is no contribution to 
${\partial \Psi \over \partial \chi}$ from the integral over 
$0 < \xi < 1$.  We may write therefore
\begin{equation}
{\partial \Psi \over \partial \vec{\chi}}(\vec{\chi}) = 
\hat{\chi}~{\pi^2 \over 8}~{{\cal M}(\chi) \over \chi^2}
\label{forc}
\end{equation}
where
\begin{equation}
{\cal M} (\chi) = G \left({t_0 \over R_0}\right)^3 {dM \over dt}(t_0)~
{8 \over \pi^2}~\int_1^\infty {d \xi \over \xi ^{3\beta-1}} 
\Theta\left(\chi - {\lambda(\xi) \over \xi^\beta}\right)~~~~\ .
\label{mass}
\end{equation}
The factor ${\pi^2 \over 8}$ is introduced so that 
${\cal M}(1)$ = 1 [see Eqs.~(\ref{actm}) and (\ref{rmass}) 
below]. The equation of motion is
\begin{equation}
{d^2 \lambda \over d \xi^2} = 
- {\pi^2 \over 8}~{\xi^{3\beta-2} \over \lambda^2}~ 
{\cal M}\left(\chi = {\lambda \over \xi^\beta}\right)~~\ .
\label{speom}
\end{equation}
Equations (\ref{mass}) and (\ref{speom}) were first obtained by 
Fillmore and Goldreich \cite{FG}.  In terms of the parameter 
$\epsilon$ which these authors introduced, 
$\beta = {2 \over 3} + {2 \over 9\epsilon}$.  Bertschinger \cite{B}
discussed the case $\epsilon = 1~(\beta = {8 \over 9})$.  

Following these authors, consider a spherically symmetric overdensity 
in an otherwise homogeneous, flat and purely matter dominated universe.  
Assume the overdensity has a power law profile
\begin{equation}
{\delta M_i \over M_i} \propto \left({1 \over M_i}\right)^\epsilon
\label{FGprof}
\end{equation}
where $M_i$ is the mass interior to radius $r_i$ in the homogeneous 
universe, and $\delta M_i$ is the corresponding overdensity.  The 
exponent $\epsilon$ must be in the range $0 < \epsilon \leq 1$. 
$\epsilon = 1$ describes the case where the overdensity is a 
point mass.  It can be shown \cite{FG,B,STW} that the time 
evolution of such an overdensity is self-similar with scale 
factor $R(t) \propto t^{{2 \over 3} + {2 \over 9\epsilon}}$.  
The mass interior to the turnaround radius $R(t)$ is 
\begin{equation}
M(t) = {\pi^2 \over 8}~{R(t)^3 \over G~t^2}~~~~\ .
\label{actm}
\end{equation}
Note that $M(t) \propto t^{2 \over 3 \epsilon}$, consistent 
with Eq.~(\ref{prop}).  Therefore, Eq.~(\ref{mass}) may be 
rewritten
\begin{equation}
{\cal M} (\chi) = {2 \over 3 \epsilon}
~\int_1^\infty {d \xi \over \xi ^{1 + {2 \over 3 \epsilon}}}
\Theta\left(\chi - 
{\lambda(\xi) \over \xi^{{2 \over 3} + {2 \over 9 \epsilon}}}\right)~~~\ .
\label{rmass}
\end{equation}
We have ${\cal M}(1) = 1$ since the argument of the $\Theta$-function 
in Eq.~(\ref{rmass}) is positive for all $\xi$ when $\chi = 1$.  
Thus 
\begin{equation}
M(r,t) = M(t) {\cal M}\left({r \over R(t)}\right)
\label{intm}
\end{equation}
is the mass interior to radius $r$ at time $t$.

\subsection{The axially symmetric case}

In the axially symmetric case, the model is specified by 
giving $\beta$, ${d M \over d\Omega d\tau}|_0(\alpha)$
and the initial velocity field 
\begin{equation}
\vec{\nu}_{\rm in} (\theta, \varphi) = \hat{r}~\nu_r(\alpha) + 
\hat{\theta}~\nu_\theta(\alpha) + \hat{\varphi}~\nu_\varphi(\alpha)~~~\ ,
\label{axsinv}
\end{equation}
where $\alpha \equiv {\pi \over 2} - \theta$.  If reflection 
symmetry is assumed in addition to axial symmetry, 
${d M \over d\Omega d\tau}|_0$, $\nu_r$ and $\nu_\varphi$ 
are even under $\alpha \rightarrow - \alpha$, whereas 
$\nu_\theta$ is odd.  One degree of freedom is removed 
from $\vec{\nu}_{\rm in}$ by the choice of $R$.  For example, 
we may choose $R$ such that all the particles in the equatorial 
plane are at first turnaround, in which case $\nu_r (0) = 0$. 

In view of the symmetry, we have
\begin{equation}
\vec{\lambda}(\theta,\varphi;\xi) = \hat{z} \lambda_z(\alpha,\xi) +
\lambda_\rho(\alpha,\xi)[\hat{x} \cos\gamma(\alpha,\varphi;\xi) + 
\hat{y} \sin\gamma(\alpha,\varphi;\xi)]~~~\ ,
\label{axsc}
\end{equation}
where the $\varphi$ dependence of $\gamma$ is trivial, i.e.
\begin{equation}
\gamma(\alpha,\varphi;\xi) = \varphi + \Delta(\alpha,\xi)~~~\ .
\label{triv}
\end{equation}
The $z$ component of angular momentum is conserved:
\begin{equation}
j_z(\alpha) = 
\lambda_\rho(\alpha,\xi)^2 {\partial\Delta \over \partial\xi}
(\alpha,\xi) = 
\nu_\varphi(\alpha)~\cos(\alpha)~~~~\ .
\label{lzcon}
\end{equation}
$\Delta(\alpha, \xi)$ is determined by solving Eq.~(\ref{lzcon}) 
with the initial condition $\Delta(\alpha, 0) = 0$.  The equations 
of motion for the other degrees of freedom are
\begin{eqnarray}
{\partial^2 \lambda_z \over \partial \xi^2} &=& 
- {1 \over \xi^{\beta-2}}~ 
{\partial\Psi \over \partial \chi_z}
\left(\vec{\chi} = {\vec{\lambda} \over \xi^\beta}\right)\nonumber\\
{\partial^2 \lambda_\rho \over \partial \xi^2} &=&
- {1 \over \xi^{\beta-2}}~
{\partial\Psi \over \partial \chi_\rho}
\left(\vec{\chi} = {\vec{\lambda} \over \xi^\beta}\right) +
{j_z(\alpha)^2 \over \lambda_\rho^3}~~~\ ,
\label{axeom}
\end{eqnarray}
with initial conditions
\begin{eqnarray}
\lambda_z(\alpha,1) &=& \sin\alpha~~,~~
\lambda_\rho(\alpha,1) = \cos\alpha\nonumber\\
{\partial \lambda_z \over \partial \xi}(\alpha,1) &=&
\sin\alpha~\nu_r(\alpha) - \cos\alpha~\nu_\theta(\alpha)\nonumber\\
{\partial \lambda_\rho \over \partial \xi}(\alpha,1) &=&
\cos\alpha~\nu_r(\alpha) + \sin\alpha~\nu_\theta(\alpha)~~~\ .
\label{axinc}
\end{eqnarray}

\section{The Model}

\subsection{Assumptions and general properties}

To first approximation when discussing its large scale 
properties, the model is spherically symmetric. So we 
start with the spherically symmetric self-similar 
infall model \cite{FG,B} introduced in Section II.D.  
The dimensionless properties of that model are determined 
in terms of a single parameter $\epsilon = {2 \over 9 \beta - 6}$.  
In an average sense, $\epsilon$ is related to the slope of 
the power spectrum of matter density perturbations \cite{2pt}.  
On galactic scales (Mpc), $\epsilon$ is of order 0.3~\cite{STW}.

In the spherically symmetric model, the angular momentum 
of each particle vanishes and all orbits are radial.  The 
particle positions are given by 
\begin{equation}
\vec{r}(\theta, \varphi, \tau; t) = 
R(\tau)~\hat{n}(\theta, \varphi)~\lambda\left({t \over \tau}\right)~~~\ ,
\label{shpspp}
\end{equation}
where $R(t)$ is the turnaround radius at time $t$, and $\tau$ 
is the time when the particle reached first turnaround.  The 
function $\lambda(\xi = {t \over \tau})$ is obtained by 
solving Eqs.~(\ref{mass}) and (\ref{speom}).  This can be 
done numerically on even a modest computer.  Fig.~\ref{lambda} 
shows the function $\lambda(\xi)$ for $\epsilon = 0.3$. 
Fig.~\ref{selfsim} shows the corresponding phase space 
distribution.

To apply the description to an actual halo, we must 
provide {\it two} dimensionful properties of the halo.
Two of the best measured properties of the Milky Way halo
are the rotation velocity at the position of the Sun 
$v_{\rm rot}(r_\odot = 8.5~{\rm kpc}$) = 220 km/s, and the 
age $t_0$ = 13.7 Gyr.  We will ignore the uncertainties on 
these two numbers.  

Fig.~\ref{rotcurv} shows the dimensionless rotation velocity 
squared $\nu^2(\epsilon, \chi)$ as a function of dimensionless
galactocentric distance $\chi = {r \over R}$ for several 
values of $\epsilon$.  $\nu(\epsilon, \chi)$ is defined by 
\begin{equation}
v_{\rm rot}^2 (r) = {G M(r) \over r} = 
\nu^2(\epsilon, {r \over R}) {G M \over R}~~~\ ,
\label{nu}
\end{equation}
where $M(r)$ is the mass interior to $r$, and $M = M(R)$.
Therefore
\begin{equation}
\nu^2(\epsilon, \chi) = {M(r) \over M} {R \over r} = 
{\cal M(\epsilon, \chi) \over \chi}~~~~\ .
\label{nu2}
\end{equation}
${\cal M}(\epsilon, \chi)$ is given by Eq.~(\ref{rmass}).  
Note that $\nu(\epsilon, 1) = 1$.  Combining Eqs.~(\ref{actm})
and (\ref{nu}), we have
\begin{equation}
R = {\sqrt{8} \over \pi} v_{\rm rot}(r)~t~
\nu^{-1}(\epsilon,{r \over R})~~~\ .
\label{detR}
\end{equation}
The turnaround radius $R$ is obtained by solving Eqs.~(\ref{detR}) 
with $t$ = 13.7 Gyr and $v_{\rm rot}(r_\odot)$ = 220 km/s.  
Table \ref{tbl1} gives $R$ and $M$ for various values of 
$\epsilon$.  This is an update of Table \ref{tbl1} in 
ref.~\cite{STW}.  

The model is characterized by a set of outer caustics located where 
the outflows reach their maximum radii $R_n$ ($n = 1, 2, 3 ...$)
before falling back in.  The properties of the outer caustics
will be discussed in detail in Section III.B.  In addition to 
these outer caustics, a galactic halo necessarily has inner 
caustics \cite{inner}.  In the spherically symmetric model, 
the inner caustics have all collapsed to the central point.  
Indeed, since its angular momentum vanishes, each particle 
passes through the center in the course of each in and out 
fall.  The center is then a singular point, where the density 
diverges.  Fig.~\ref{rotcurv} shows that, in the spherically 
symmetric model, the halo contribution to the rotation curve 
is constant near $r = 0$ for $\epsilon \sim$ 0.3, implying that 
the density $d(r) \propto {1 \over r^2}$ as $r \rightarrow 0$.

The spherically symmetric model does not adequately describe 
the inner parts of the halo.  In actual galaxies, the central
parts are dominated by baryons and the halo contribution to the 
rotation curve goes to zero as $r \rightarrow 0$.  As pointed out 
in ref.~\cite{STW}, the depletion of the inner halo, as compared 
to the prediction of the spherically symmetric self-similar model, 
is neatly accounted for by the angular momentum that the dark matter
particles are expected to carry because angular momentum causes the 
dark matter particles to avoid the center.  Indeed, by constraining 
the distance scale (effective core radius) over which the inner halo 
is depleted, one can estimate the average specific angular momentum 
of the halo particles \cite{STW}.

When angular momentum is introduced, the inner caustics spread 
over surfaces with cusps.  (The word ``cusp" has in this paper 
the same meaning as in Catastrophe Theory, unrelated to the notion 
of ``cuspy halos".)  In ref.~\cite{inner}, the catastrophe structure 
of the inner caustics was investigated as a function of the angular
momentum distribution of the infalling particles.  If the angular 
momentum distribution is dominated by net overall rotation, the 
inner caustic is a ring (i.e. a closed tube) whose cross-section 
is a section of the elliptic umbilic ($D_{-4}$) catastrophe \cite{sing}.  
We call this type of inner caustic a ``caustic ring of dark matter" 
\cite{crdm,sing}.  Figs.~\ref{flow} and \ref{dimen} show such a 
ring in cross-section. Fig.~\ref{flow} shows the flow of dark 
matter in the neighborhood of the caustic, whereas Fig.~\ref{dimen} 
shows the definition of its radius $a$ and transverse sizes $p$ and 
$q$.  Caustic rings of dark matter are described in more detail in 
Section III.C below.  When the initial velocity distribution is 
dominated by a curl-free (non rotational) component, the inner 
caustic has a ``tent-like" structure, which may be described 
qualitatively as a caustic ring of dark matter whose (inner) 
radius $a$ shrunk to zero while its outer radius $a+p$ was 
held fixed.  See ref.~\cite{inner} for a detailed description.

As summarized in Section IV, there is evidence for caustic 
rings of dark matter in the Milky Way and other spiral galaxies.  
The evidence is for caustic rings, as opposed to the tent-like 
inner caustics that occur when the angular momentum distribution 
is irrotational.  Hence we assume that the angular momentum 
distribution of the infalling dark matter is dominated by net 
overall rotation.  We assume furthermore that it is axially 
and reflection symmetric, and consistent with self-similarity.
To our previous spherically symmetric description depending on 
one dimensionless parameter $\epsilon$, we add therefore the 
dimensionless initial velocity field of Eq.~(\ref{axsinv}).  
The specific angular momentum of the particle at turnaround 
radius $R(\tau)$ and angular coordinates $(\theta, \varphi)$ 
is then 
\begin{equation}
\vec{l}(\theta, \varphi, \tau) = 
{R(\tau)^2 \over \tau} \vec{j}(\theta, \varphi)
\label{spam}
\end{equation}
where
\begin{equation}
\vec{j}(\theta, \varphi) = 
\hat{n}(\theta, \varphi) \times \vec{\nu}_{\rm in}(\theta, \varphi)
= - \nu_\varphi(\alpha) \hat{\theta} +
\nu_\theta(\alpha) \hat{\varphi}~~~\ .
\label{j}
\end{equation}
The assumption that the initial velocity distribution is dominated 
by net overall rotation means $\nu_\varphi >> |\nu_r|, |\nu_\theta|$.  
The simple case of initial rigid rotation is 
$\nu_\varphi(\alpha) = \nu_\varphi(0) \cos \alpha$, 
$\nu_r = \nu_\theta = 0$.  

It was mentioned already that, when angular momentum is added 
to the spherically symmetric infall model, the inner parts of 
the halo are depleted and the halo contribution to the rotation 
curve goes to zero when $r \rightarrow 0$, consistent with 
observation.  In the Milky Way and in many other large spiral 
galaxies, the baryon contribution to the rotation curve is such 
as to make the curve approximately flat at small $r$, and with 
approximately the same value at small $r$ as at large $r$.  In 
other words, the baryons and the dark matter ``conspire" to keep 
the rotation velocity approximately constant for all radii where 
it has been measured.  Now, recall that the rotation curve of 
the spherically symmetric self-similar model is very nearly 
constant for all $r$ when $\epsilon \sim 0.3$; see Fig.~\ref{rotcurv}.  
Following ref.\cite{STW}, we take account of the gravitational 
force exerted by the baryons in the Galaxy by using the 
potential of the spherically symmetric model to calculate 
the motion of the dark matter particles in the axially 
symmetric model.  This should be a good approximation 
when calculating the properties of low $n$ flows since 
they are made of particles falling in from distances much 
greater than the size of the baryonic disk.  Furthermore, 
the caustic rings are made of particles that move in or 
near the equatorial plane and therefore respond to the 
same gravitational forces as produce the nearly flat 
rotation curve of the Galaxy.

In ref. \cite{STW}, the average value $\bar{j}$ of 
$j(\theta, \varphi) \equiv |\vec{j} (\theta, \varphi)|$ 
was estimated to be 0.2 by requiring the halo contribution 
to the rotation curve to approach zero as $r \rightarrow 0$ 
over a distance scale (effective core radius) consistent with 
observations.  Subsequently, evidence was found for caustic 
rings of dark matter in the Milky Way \cite{crdm,milk}.  This 
evidence is consistent with the maximum value $j_{\rm max}$ 
of $j(\theta, \varphi)$ being equal to 0.263.  $j_{\rm max}$ is 
to be identified with $\nu_\varphi(0)$.  Although the precise 
relationship between $\bar{j}$ and $j_{\rm max}$ depends 
on the full $j(\theta, \varphi)$-distribution, the values 
$\bar{j} \sim 0.2$ and $j_{\rm max} \simeq 0.263$ are 
consistent with one another.  For example, for the simple 
case of initial rigid rotation, one has 
$j_{\rm max} = {4 \over \pi} \bar{j}$ and 
hence $j_{\rm max} = 0.255$ for $\bar{j} = 0.2$.  

Both $j_{\rm max} \simeq 0.263$ and $\bar{j} \sim 0.2$ 
are the outcome of fits to observations in the old 
$\Omega = \Omega_{\rm m} = 1$ cosmology with $H_0$ = 
70 km/(s Mpc) in which the age of our galaxy is 9.31 Gyr.  
Since we now adopt the age of 13.7 Gyr, while keeping the 
rotation velocity (220 km/s) unchanged, all length scales 
are stretched by the factor 13.7/9.31= 1.47.  Because the 
effective core radius is approximately proportional to 
$\bar{j}$ and the caustic ring radii approximately proportional 
to $j_{\rm max}$, the values of these parameters should be 
reduced by the factor 1.47 compared to the old $\Omega = 
\Omega_{\rm m} = 1$ model.  So the fit to observations 
yields $\bar{j} \sim 0.14$ and $j_{\rm max} \simeq 0.179$ 
in Concordance Cosmology.  More precisely, because the 
caustic ring radii are not exactly proportional to 
$j_{\rm max}$, the fit of the model to the evidence 
in Concordance Cosmology (see Section IV.D) yields 
$j_{\rm max}$ = 0.186, which is the value adopted 
in this paper.  To avoid confusion, we write 
henceforth $j_{\rm max,old}$ for values in the 
old $\Omega = \Omega_{\rm m} = 1$ model, and 
simply $j_{\rm max}$ for values in the Concordance 
Cosmology.  

For the sake of definiteness, the model presented here
assumes $\epsilon$ = 0.3.

\subsection{Outer caustic spheres}

Caustics are at the boundaries between regions in physical 
space with differing number of flows.  On one side of a 
caustic there are two more flows than on the other.  When 
an isolated halo is approached from the outside, the local 
number of flows increases.  First, there is one flow, then 
three, then five, and so on.  The boundary between the region 
with one flow and the region with three flows is the location 
of the first outer caustic.  Likewise, the second outer caustic 
is at the boundary between the region with three flows and the 
region with five flows, and so on.  See Fig. 1 for an illustration.  

The outer caustics are simple fold ($A_2$) castastrophes located 
on topological 2-spheres surrounding the galaxy.  The density in 
the 2 extra flows on the inside of an outer caustic diverges on 
the 2-sphere, with the following characteristic behaviour:
\begin{equation}
d(x) = {A \over \sqrt{x}} \Theta(x) [1 + {\cal O}(x)]
\label{sqrt}
\end{equation}
where $x$ is the distance to the 2-sphere, measured positively 
on the inside, $\Theta(x)$ is the step function, and $A$ is a 
constant which we call the fold coefficient.  In the absence 
of symmetry, $A$ varies with location on the 2-sphere.  The 
divergence of $d$ as $x \rightarrow 0_+$ is cut off by the 
velocity dispersion of the dark matter particles.  

Let us emphasize that the {\it existence} of outer 
caustics with the properties listed in the previous 
paragraph follows exclusively from the existence of cold
collisionless dark matter.  

\subsubsection{Radii $R_n$}

Because angular momentum has little influence on the outer 
caustics, we use the spherically symmetric model to describe 
their properties and, in particular, to estimate the galactocentric 
radii $R_n$ where they occur and their fold coefficients $A_n$.  
As before, each spherical shell is labeled by the time $\tau$ 
when it was at its first turnaround.  At time $t$, the radius 
of a shell is 
\begin{equation}
r(\tau, t) = R(\tau) \lambda\left({t \over \tau}\right)
= R(t) {1 \over \xi^\beta}
\lambda(\xi)\bigg|_{\xi = {t \over \tau}}~~~~\ ,
\label{shrad}
\end{equation}
where we used $R(t) \propto t^\beta$.  The outer caustics are at 
radii
\begin{equation}
R_n = R(t) \Lambda(\xi_n)
\label{Rn}
\end{equation}
where the $\xi_n$ are the locations of the maxima
of $\Lambda(\xi) \equiv \xi^{-\beta} \lambda(\xi)$.

\subsubsection{Fold coefficients $A_n$}

From Eqs.~(\ref{rmass}) and (\ref{intm}), one may derive 
the following formula for the density $d(r,t)$ as a sum 
over flows 
\begin{equation}
d(r,t) = \sum_j d_j(r,t) = 
{1 \over 6 \pi \epsilon} {M(t) \over R(t) r^2}
\sum_j {1 \over \xi_j(r,t)^{1 + {2 \over 3 \epsilon}}
\bigg|{d \Lambda \over d \xi} (\xi_j(r,t))\bigg|}
\label{dsum}
\end{equation}
where the $\xi_j(r,t)$ are the solutions of $r = R(t)~\Lambda(\xi)$.
Near a maximum $\xi_n$ of $\Lambda (\xi)$, we have 
\begin{eqnarray}
\Lambda(\xi) &=& \Lambda(\xi_n) + 
{1 \over 2} {d^2 \Lambda \over d \xi^2}(\xi_n) (\xi - \xi_n)^2
+ {\cal O}(\xi - \xi_n)^3\nonumber\\
{d \Lambda \over d \xi} &=& + 
{d^2 \Lambda \over d \xi^2}(\xi_n) (\xi - \xi_n) + 
{\cal O}(\xi - \xi_n)^2~~~~\ ,
\label{Taylor}
\end{eqnarray}
with ${d^2 \Lambda \over d \xi^2}(\xi_n) < 0$.  Let us label 
$j = (n+)$ and $(n-)$ the two flows which form the caustic at 
$r = R_n$.  Then (the $t$ dependence is suppressed in the
remainder of this subsection) 
\begin{equation}
\xi_{n\pm}(r) - \xi_n = 
\pm \sqrt{2 (R_n - r) \over - R {d^2 \Lambda \over d \xi^2}(\xi_n)}
[1 + {\cal O}(\xi_{n\pm} - \xi_n)]
\label{xinpm}
\end{equation}
for $r$ close to, but less than, $R_n$.  Combining Eqs.~(\ref{dsum}) 
and (\ref{xinpm}), we find 
\begin{equation}
d_{n\pm}(r) = {1 \over 2} {A_n \over \sqrt{R_n - r}}
\Theta(R_n - r) [1 + {\cal O}(\sqrt{R_n - r})]
\label{dnpm}
\end{equation}
with 
\begin{equation}
A_n = {1 \over 3 \pi \epsilon} {M \over \sqrt{2 R} R_n^2}
{1 \over \xi_n^{1 + {2 \over 3 \epsilon}} 
\sqrt{- {d^2 \Lambda \over d \xi^2}(\xi_n)}}~~~~~\ .
\label{An}
\end{equation}
The ${\cal O}(\sqrt{R_n - r})$ remainder in Eq.~(\ref{dnpm}) 
has opposite sign for $d_{n+}(r)$ and $d_{n-}(r)$ so that
\begin{equation}
d_n(r) = d_{n+}(r) + d_{n-}(r) = 
{A_n \over \sqrt{R_n - r}} \Theta(R_n - r) [1 + {\cal O}(R_n -r)]~~~\ .
\label{dnc}
\end{equation}
Table \ref{tbl2} lists the present $R_n$ and $A_n$ values in 
the spherically symmetric infall model with $\epsilon$ = 0.3, 
for $n$ = 1, 2, ... 20.

\subsection{Inner caustic rings}

It was mentioned already that dark matter halos necessarily 
have inner caustics, that the catastrophe structure of the 
inner caustics depends on the angular momentum distribution 
of the infalling dark matter particles \cite{inner}, and that 
the inner caustics are rings if, as we assume to be the case, 
the angular momentum distribution is dominated by net overall 
rotation.  The caustic ring singularity was described in 
ref.~\cite{sing} which the reader may wish to consult for 
background information.

In the limit of axial and reflection symmetry and where the 
transverse sizes $p$ and $q$ of a caustic ring are much smaller 
than its radius $a$, the distribution of dark matter particles 
in the vicinity of the caustic ring is given by the particle 
positions
\begin{eqnarray}
z(\alpha, \eta) &=& b \alpha \eta  \nonumber\\
\rho(\alpha, \eta) &=& a + {1 \over 2} u (\eta - \eta_0)^2
- {1 \over 2} s \alpha^2~~~~\ .
\label{crfl}
\end{eqnarray}
We use cylindrical coordinates $(z, \rho, \varphi)$ for position
in physical space.  Eqs.~(\ref{crfl}) give particle positions 
at a particular time $t$, which is not shown explicitly.  The 
particles are labeled by parameters $(\alpha, \eta)$.  As before, 
$\alpha \equiv {\pi \over 2} - \theta$ where $\theta$ is the polar 
angle of the particle at the time of its first turnaround.  $\eta$ 
is the time when the particle crosses the $z = 0$ plane in the 
course of its flow through the caustic ring.  $t - \eta$ can 
be thought of as the age of the particle. The particles labeled 
$(\alpha, \eta)$ form a circle of radius $\rho(\alpha, \eta)$ 
at a height $z(\alpha, \eta)$ above the $z = 0$ plane.  

Fig.~\ref{flow} plots $(\rho(\alpha, \eta), z(\alpha, \eta))$ 
for continuous $\eta$, and discrete values of $\alpha$.  The 
lines in Fig.~\ref{flow} are the trajectories of the particles 
forming the flow, except that positions are plotted as a function 
of age, whereas for ordinary trajectories position is plotted as 
a function of time.  Let us call the lines of Fig.~\ref{flow} 
``age trajectories''.  Fig.~\ref{flow} shows that particle 
density diverges on a closed line which has the shape of a 
isosceles triangle, but with cusps instead of angles.  We call 
that shape a``tricusp''.  The location of the tricusp is shown 
in Fig.~\ref{dimen} for the flow of Fig.~\ref{flow}.  It is 
the envelope of the age trajectories.  There are four flows 
everywhere inside the tricusp and two flows everywhere outside.  
The caustic, i.e. the surface where the density diverges, lies 
at the boundary between the region with four flows and the region 
with two flows.

The physical space density is given by \cite{sing}
\begin{equation}
d(\rho, z) = {1 \over \rho} \sum_{j=1}^{N(\rho,z)}
{dM \over d\Omega d\eta}(\alpha,\eta)
{\cos\alpha \over |D(\alpha, \eta)|}
\bigg|_{(\alpha_j(\rho,z), \eta_j(\rho,z))}
\label{denax}
\end{equation}
where $\alpha_j(\rho,z)$ and $\eta_j(\rho,z)$, with
$j = 1~...~N(\rho,z)$, are the solutions of $\rho(\alpha,\eta) = \rho$
and $z(\alpha,\eta) = z$. $N(\rho,z)$ is the number of flows at
position $(\rho,z)$; thus, $N=4$ inside the tricusp and $N=2$
outside.  $D(\alpha,\eta)$ is the Jacobian determinant of the
map $(\alpha,\eta) \rightarrow (\rho,z)$:
\begin{equation}
D(\alpha,\eta) \equiv
\det\left({\partial(\rho,z) \over \partial (\alpha,\eta)}\right)
= - b [u \eta (\eta- \eta_0) + s \alpha^2]~~~\ .
\label{D}
\end{equation}
${dM \over d\Omega d\eta} = {dM \over 2 \pi \cos\alpha d\alpha d\eta}$
is the mass falling in per unit solid angle and unit time.  The tricusp
perimeter is the locus of points $(\rho(\alpha, \eta), z(\alpha, \eta))$
for which $D(\alpha, \eta) = 0$.  We call $p$ and $q$ the sizes of the
tricusp in the $\rho$ and $z$ directions respectively; see
Fig.~\ref{dimen}.
They are given by
\begin{equation}
p = {1 \over 2} u \eta_0^2~~,~~~
q = {\sqrt{27} \over 4} {b \over \sqrt{us}}~p~~~\ .
\label{pq}
\end{equation}
Let us write the velocity of particle labeled $(\alpha, \eta)$ as
$\vec{v}(\alpha, \eta) =
v_\varphi (\alpha, \eta) \hat{\varphi} + v_\rho (\alpha, \eta) \hat{\rho}
+ v_z (\alpha, \eta) \hat{z}$.  The main component of velocity
is in the $\hat{\varphi}$ direction: $v_\varphi \simeq v$.  In the case 
of a stationary flow, the velocity components in the $\hat{\rho}$ 
and $\hat{z}$ directions are
\begin{equation}
v_\rho = - {\partial \rho \over \partial \eta} = - u (\eta - \eta_0)~~,~~~
v_z = - {\partial z \over \partial \eta} = - b \alpha~~~\ .
\label{tranvel}
\end{equation}
Here we use the fact that, in case of stationary flow, the
particle positions are functions only of their age $t - \eta$.
Caustic rings grow in mass and radius on cosmological time scales.
Therefore stationarity is not an exact property of caustic rings. 
However, it is an excellent approximation.  In the self-similar 
model the growth in physical size is tantamount to expansion in 
all directions by the scale factor $R(t)$.  Hence $v_\rho$ 
in Eq.~(\ref{tranvel}) should be replaced by 
$v_\rho + {\dot{R} \over R} \rho$ and $v_z$ by 
$v_z + {\dot{R} \over R} z$.  However the corrections are small, 
of order 1\% for $n=1$ and less for $n \geq 2$, and will be ignored 
henceforth.  For stationary flow, the speed $v$ is related to $u$ by
\begin{equation}
u = {v^2 \over a}
\label{ca}
\end{equation}
since $u$ is the centrifugal acceleration of the particles at
$(z, \rho) = (0, a)$.

Our description of a caustic ring in the limit of axial and
reflection symmetry, and where the transverse sizes $p$ and $q$ 
of the ring are much smaller than its radius $a$, is in terms of 
six parameters: $a,~b,~u,~\eta_0,~s$ and ${dM \over d\Omega d\eta}$.  
We now turn to the self-similar infall model to obtain predictions 
for many of these parameters.  As was discussed in Section III.A, 
the particles are assumed to fall in the gravitational potential 
produced by the mass distribution $M(r,t)$ of the {\it spherically 
symmetric} self-similar infall model.  The equation of motion is 
obtained therefore by combining Eqs.~(\ref{eom}) and (\ref{forc}):
\begin{equation}
{\partial^2 \vec{\lambda} \over \partial \xi^2} =
- \hat{\lambda}~{\pi^2 \over 8} \xi^{3\beta-2}
{1 \over \lambda^2} {\cal M}\left({\lambda \over \xi^\beta}\right)~~~\ ,
\label{eom3}
\end{equation}
and using for ${\cal M}(\chi)$ the solution of Eqs.~(\ref{mass}) and 
(\ref{speom}).  The initial conditions are 
\begin{equation}
\vec{\lambda}(\theta, \varphi; 1) = \hat{n}(\theta, \varphi)~~~,~~~
{\partial \vec{\lambda} \over \partial \xi} (\theta, \varphi; 1) = 
\vec{\nu}_{\rm in}(\theta, \varphi)
\label{axincon}
\end{equation}
where $\vec{\nu}_{\rm in} (\theta, \varphi)$ is the initial velocity 
defined in Eq.~(\ref{axsinv}).

\subsubsection{Radii $a_n$}

The radius $a_n$ of the $n^{\rm th}$ caustic ring is the distance 
of closest approach to the galactic center of the particles in the 
equatorial plane for the $n^{\rm th}$ in and out flow.  To obtain 
the $a_n$ we solve
Eq.~(\ref{eom3}) with $\alpha = \theta - {\pi \over 2} = 0$. By 
reflection symmetry, $\nu_\theta(0) = 0$.  Also, $\nu_r(0) = 0$ by
definition of the turnaround radius $R$.  Eq.~(\ref{eom3}) implies
conservation of angular momentum.  We have
\begin{equation}
\lambda^2(\xi)~{d{\varphi} \over d \xi} = \nu_\varphi(0)
\label{amcon}
\end{equation}
for the particles in the equatorial plane.  The quantity 
$\nu_\varphi(0)$ was called $j_{\rm max}$ in previous 
publications.  For the sake of consistency, we continue 
to use this name here.  The equation of motion for the 
radial degree of freedom is then
\begin{equation}
{d^2 \lambda \over d \xi^2} = + {j_{\rm max}^2 \over \lambda^3}
- {\pi^2 \over 8} \xi^{3\beta-2} 
{1 \over \lambda^2} {\cal M}\left({\lambda \over \xi^\beta}\right)~~~\ .
\label{eomr}
\end{equation}
The radial coordinate of the particles that reached first turnaround 
at time $\tau$ is 
\begin{equation}
r(\tau, t) = R(\tau) \lambda\left(j_{\rm max}, {t \over \tau}\right) =
R(t) {1 \over \xi^\beta} 
\lambda(j_{\rm max}, \xi)\bigg|_{\xi = {t \over \tau}}
\label{again}
\end{equation}
where $\lambda(j_{\rm max}, \xi)$ is the solution of Eq.~(\ref{eomr}) 
with initial conditions: $\lambda(j_{\rm max}, 1) = 1,~
{d \lambda \over d \xi}(j_{\rm max}, 1) = 0$.  The caustic ring radii 
$a_n$ are 
\begin{equation} a_n = R(t) \Lambda(j_{\rm max}, \xi_n^\prime) 
\label{crr}
\end{equation} 
where the $\xi_n^\prime$ are the locations of the minima of 
$\Lambda(j_{\rm max},\xi) \equiv \xi^{-\beta} \lambda(j_{\rm max},\xi)$.  
The second column of Table \ref{tbl3} lists the $a_n$ for $n$ = 
1, 2, ... 20, in the $\epsilon$ = 0.3 model with $j_{\rm max}$ = 
0.186. This value of $j_{\rm max}$ in the Concordance Cosmology was 
determined from a fit of the $a_n$ to a set of rises in the Milky 
Way rotation curve, as discussed in Section IV.

\subsubsection{Accelerations $u_n$}

The particles at $(z, \rho) = (0, a_n)$ are moving in the 
$\hat{\varphi}$ direction with speed $v_n = {l_n \over a_n}$
where 
\begin{equation}
l_n = j_{\rm max} {R^2(\tau_n) \over \tau_n} =
j_{\rm max} {R^2(t) \over t} {1 \over \xi_n^{\prime~2\beta-1}}~~~\ ,
\label{ln}
\end{equation}
is their specific angular momentum.  Combining Eqs.~(\ref{crr}) 
and (\ref{ln}), we obtain
\begin{equation}
v_n = {R(t) \over t} j_{\rm max} {1 \over 
\xi_n^{\prime~\beta-1}~\lambda(j_{\rm max}, \xi_n^\prime)}~~~\ . 
\label{vn}
\end{equation}
The $u_n$ parameters are then  
\begin{equation}
u_n = {v_n^2 \over a_n}~~~~~\ .
\label{un}
\end{equation}
The speeds $v_n$ are listed in the third column of 
Table \ref{tbl3} for $\epsilon = 0.3$ and $j_{\rm max}$ = 0.186.  
Note that the $v_n$ only have a weak dependence on $j_{\rm max}$ 
because the denominator in the RHS of Eq.~(\ref{vn}) is 
proportional to $j_{\rm max}$ in the small $j_{\rm max}$ limit.  

\subsubsection{Infall rates~ 
${d M \over d \Omega d \eta}\bigg|_n$}

The infall rates are obtained by noting that 
$M(t) \propto t^{2 \over 3 \epsilon}$ and assuming that the 
infall is isotropic.  We have 
\begin{equation}
{d M \over d \Omega d \eta} \bigg|_n = 
{d M \over d \Omega d \tau} |{d \tau \over d \eta}| \bigg|_n =
{1 \over 4 \pi} {2 \over 3 \epsilon} {M(\tau_n) \over \tau_n}
{\tau_n \over t} 
= {1 \over 6 \pi \epsilon} {M(t) \over t} 
{1 \over \xi_n^{\prime {2 \over 3\epsilon}}}~~~~~\ .
\label{infrn}
\end{equation}
That ${d \tau \over d \eta}|_n = {d \tau \over dt}|_n = {\tau_n \over t}$
follows from self-similarity since, in the absence of any time scale, 
$\tau_n$ and $t$ must be proportional to one another.  The fourth column
of Table \ref{tbl3} shows the ${d M \over d \Omega d \eta}\bigg|_n$ 
values in the model.

\subsubsection{$b_n,~\eta_{0,n},~s_n$}

The remaining parameters $b_n$, $\eta_{0,n}$ and $s_n$
are also predicted by the self-similar infall model 
but are related to more detailed properties of $\vec{\nu}_{\rm in}$ 
near $\alpha = 0$.  We have \cite{sing}
\begin{equation}
b_n = v_n [\cos \delta_n(0) + \phi^\prime(0) \sin \delta_n(0)]
\label{bn}
\end{equation}
and 
\begin{equation}
\eta_{0,n} = {a_n \over v_n}~ 
{\phi^\prime(0) \cos \delta_n(0) - \sin \delta_n(0) \over 
\phi^\prime(0) \sin \delta_n(0) + \cos \delta_n(0)}~~~~\ ,
\label{en}
\end{equation}
where 
\begin{equation}
\phi^\prime(0) = - {1 \over j_{\rm max}} 
{d \nu_\theta \over d \alpha}(0)
\label{phio}
\end{equation}
and 
\begin{equation}
\delta_n (0) = - {\pi \over 2} + 
j_{\rm max} \int_1^{\xi_n^\prime} {d \xi \over \lambda (\xi)^2}~~~\ .
\label{dno}
\end{equation}
Ref. \cite{sing} gives also a formula for the $s_n$ involving higher 
derivatives of $\nu_\theta(\alpha)$ and $\nu_\varphi(\alpha)$ at
$\alpha = 0$.  Generally, $s_n$ and $a_n$ are of the same order 
of magnitude.

\subsection{Interpolating between inner and outer caustics}

The model predicts the full phase space distribution of the halo
at all times, i.e. it predicts the number $N(\vec{r},t)$ of flows 
at all locations, the flow densities $d_j(\vec{r},t)$ and the flow 
velocities $\vec{v}_j(\vec{r},t)$.  To obtain these quantities 
one uses, in principle, the following procedure.  Obtain
$\vec{\lambda}(\theta,\varphi;\xi)$ by solving Eqs.~(\ref{eom3})
with initial conditions (\ref{axincon}) for all $(\theta,\varphi)$.
Then find the solutions $(\theta,\varphi,\tau)_j$, $j = 1,2 ... N$, of
\begin{equation}
\vec{r} = R(t) \left({\tau \over t}\right)^\beta
\vec{\lambda}(\theta, \varphi; {t \over \tau})~~~\ .
\label{solj}
\end{equation}
The number of solutions $N$ is the number of flows at location 
$\vec{r}$ at time $t$.  The velocities of the flows are 
\begin{equation}
\vec{v}_j(\vec{r}, t) = {R(t) \over t} 
\left({\tau_j \over t}\right)^{\beta-1}
{\partial \vec{\lambda} \over \partial \xi}
(\theta_j, \varphi_j; \xi = {t \over \tau_j})~~~~\ .
\label{vecj}
\end{equation}
Their densities are 
\begin{equation}
d_j(\vec{r}, t) = {d M \over d \Omega d \tau}(\tau)
{\sin \theta \over 
\bigg|\det \bigg({\partial (x,y,z) \over 
\partial (\theta, \varphi, \tau)} \bigg)\bigg|} 
\Bigg|_{(\theta_j, \varphi_j, \tau_j)}~~~\ .
\label{dj}
\end{equation} 
It is certainly possible to carry out this procedure in 
practice, but it would be laborious.  We note that the 
underlying formalism keeps track of all particles at all times.  
Since the particles are indistinguishable, this is far more 
information than the full phase space distribution at all times.  
Many practical tests of the model are sensitive only to the phase 
space distribution.

The purpose of this section is to give approximate formulas
for the full phase space distribution today, based upon the 
smooth interpolation between inner and outer caustics. The 
generalization to other times is straightforward since all 
lengths scale as $t^\beta$, all velocities as $t^{\beta-1}$ 
and all densities as $t^{-2}$.

At an arbitrary location $\vec{r}$ which is not within 
the tricusp tube of a caustic ring, the number of flows 
is $N(r) = 1 + 2 \sum_n \Theta(R_n - r)$ where $\Theta(x)$ 
is the step-function.  For example at $r =$ 130 kpc, there 
are 9 flows because four outer caustics have radius larger 
than 130 kpc; see Table \ref{tbl2}.  If $\vec{r}$ is within 
the tricusp tube of a caustic ring, the number of flows is 
$N(r) = 3 + 2 \sum_n \Theta(R_n - r)$ because there are 
two more flows inside the tricusp tube of a caustic ring 
than outside.

For all $r$ the flow speeds $v_{n\pm}(r)$ can be read off 
from Fig.~\ref{selfsim}.  This neglects the effect of angular 
momentum on flow speed, a small effect.  The flow speeds
can also be estimated more crudely by using the formula
\begin{equation}
v_{n\pm}(r) = 
v_{\rm rot} \sqrt{2 \ln \left({R_n \over r}\right)}~~\ .
\label{vnr}
\end{equation}
Eq.~(\ref{vnr}) would follow from energy conservation in 
the static gravitational potential 
$\Phi(r) = v_{\rm rot}^2 \ln \left({1 \over r}\right)$.
Because the actual potential is time-dependent, energy 
conservation is only approximate.  It is least reliable 
for $n=1$ since the gravitational potential is changing 
appreciably on the time scale of the first infall.  

To estimate the density of the $n^{\rm th}$ in and out 
flows one may use 
\begin{equation}
d_{n\pm}(r) = {1 \over r^2} {1 \over v_{n\pm}(r)} 
{d M \over d \Omega d t}\bigg|_{n\pm}(r) 
\label{dnr}
\end{equation}
provided $r >> a_n$.  
Eq.~(\ref{dnr}) expresses particle number conservation in the 
spherically symmetric case.  The infall rates are given by 
\begin{equation}
{d M \over d \Omega d t}\bigg|_{n\pm}(r) =
{1 \over 6 \pi \epsilon} {M \over t}
{1 \over \xi_{n\pm}(r)^{2 \over 3 \epsilon}}
\label{infre}
\end{equation}
where the $\xi_{n\pm}(r)$ are the solutions of
$r = R(t) \Lambda(\xi)$.  Eqs.~(\ref{infre}) is obtained 
through the same steps as Eq.~(\ref{infrn}).  The infall 
rates are smooth functions of $\xi$ and may be obtained by 
interpolating between their values at the inner caustics, 
given in Table \ref{tbl3}.

In the spherically symmetric case, the flow equations 
near radius $r$ are
\begin{equation}
z(\alpha, \eta) = v(r)~\eta~\sin\alpha~~,~~
\rho(\alpha, \eta) = v(r)~|\eta|~\cos\alpha~~\ .
\label{radfl}
\end{equation}
where $v(r)$ is the speed of the flow at radius $r$.
As before, $\eta$ is the time when the particle crosses 
the $z=0$ plane.  These equations are valid for $r>>a$ 
where $a$ is the radius of the inner caustic made by that 
flow.  Close to the inner caustic, the flow is described 
by Eqs.~(\ref{crfl}).  Comparing Eqs.~(\ref{crfl}) and 
(\ref{radfl}) suggests an interpolating formula for the 
case $b=v$ and $s=a$:
\begin{equation}
z = v~\eta~\sin\alpha~~,~~ 
\rho = \sqrt{a^2 + v^2 (\eta - \eta_0)^2}~\cos\alpha~~ \ .
\label{inter}
\end{equation}
Indeed, since $u = {v^2 \over a}$, Eqs.~(\ref{inter})
reduce to Eqs.~(\ref{crfl}) for small $\eta$ and small 
$\alpha$ when $b = v$ and $s = a$, and reduce to 
Eqs.~(\ref{radfl}) for large $\eta$ and all $\alpha$.  
To accomodate $b \neq v$ and/or $s \neq a$, one may 
generalize Eqs.~(\ref{inter}) to 
\begin{equation}
z = v~\eta~\sin(g_z(\eta) \alpha)~~, ~~
\rho = \sqrt{a^2 + v^2 (\eta - \eta_0)^2} \cos(g_\rho(\eta) \alpha)
\label{inter2}
\end{equation}
where $g_z(\eta)$ and $g_\rho(\eta)$ are smooth functions which 
approach respectively ${b \over v}$ and $\sqrt{s \over a}$ when 
$\eta \rightarrow 0$, but approach one when $\eta >> \eta_0$.

Sufficiently far from a caustic ring, one may choose to neglect 
its transverse dimensions $p$ and $q$.  This is done by setting 
$\eta_0 = 0$.  In that limit, Eqs.~(\ref{inter}) may be easily 
inverted to obtain $\eta(z,\rho)$ and $\alpha(z,\rho)$.  One can 
then express the density and velocity fields directly in terms of 
position \cite{sing}:
\begin{eqnarray}
v_z &=& \mp {v \over a}
\sqrt{{1 \over 2}\left(a^2 - r^2 + 
\sqrt{(r^2 - a^2)^2 + 4 a^2 z^2}\right)}\nonumber\\
v_\rho &=& \mp {\rm sign}(z) {v \over 2 a^2 \rho} 
\sqrt{{1 \over 2}\left(r^2 - a^2 +
\sqrt{(r^2 - a^2)^2 + 4 a^2 z^2}\right)}
\left(r^2 + a^2 - \sqrt{(r^2 - a^2)^2 + 4 a^2 z^2}\right)\nonumber\\
v_\varphi &=& + \sqrt{v^2 - v_z^2 - v_\rho^2}\nonumber\\
d_+ &=& d_- = {1 \over v} {d M \over d \Omega d \tau}
{1 \over \sqrt{(r^2 - a^2)^2 + 4 a^2 z^2}}~~~\ ,
\label{exap}
\end{eqnarray}
where the $\mp$ signs are for the down and up flows.

\subsection{Adiabatic approximation}  

Except during the first couple throughfalls, the gravitational 
potential seen by a dark matter particle is slowly varying on the
oscillation time scale of its radial coordinate.  By exploiting the
adiabatic invariant for the radial motion of the particle, it is 
possible to estimate its trajectory without resorting to numerical 
integration.  This method was used in ref. \cite{FG} to show that 
the halo density $d \propto {1 \over r^2}$ as $r \rightarrow 0$ in 
the spherical infall model with $\epsilon < {2 \over 3}$.  It was 
further used in ref. \cite{STW} to obtain estimates of the densities 
and speeds of the flows at our distance from the galactic center. The 
adiabatic method gives predictions for the quantities of interest here.  
We state these predictions without proof, leaving their derivation to
a future publication.

For the outer caustics, the adiabatic method predicts
\begin{eqnarray}
\xi_n &=& \left[{4 \over \sqrt{\pi}} {4 + 3 \epsilon \over 9 \epsilon}
~n~+~1\right]^{9 \epsilon \over 4 + 3 \epsilon}\nonumber\\
R_n &=& R(t)~\xi_n^{~-{4 + 3 \epsilon \over 9 \epsilon}}
\nonumber\\
A_n &=& {2 \over 3 \pi^2 \epsilon}~{M \over R^{5 \over 2}}~
\xi_n^{~{1 \over 2}}~~~\ .
\label{adout}
\end{eqnarray}
To obtain these results, one assumes that the adiabatic invariant 
for radial motion is constant and that the potential is exactly 
logarithmic, i.e. $\Phi(r,t) = - v_{\rm rot}(t)^2 \ln(r)$.

For the inner caustics, the adiabatic method predicts
\begin{eqnarray}
\xi_n^\prime &=& \left[{4 \over \sqrt{\pi}} 
{4 + 3 \epsilon \over 9 \epsilon}
~(n - {1 \over 2})~+~1\right]^{9 \epsilon \over 4 + 3 \epsilon}\nonumber\\
a_n &=& {R(t) \over \gamma}~
(\xi_n^\prime)^{-{4 + 3 \epsilon \over 9 \epsilon}}\nonumber\\
v_n &=& {R(t) \over t}~\gamma~j_{\rm max}
\label{adin}
\end{eqnarray}
where $\gamma$ is given in terms of $j_{\rm max}$ by 
\begin{equation}
j_{\rm max} = {\pi \over 2} \sqrt{\ln (\gamma) \over \gamma^2 - 1}~~~\ .
\label{gam}
\end{equation}
To obtain Eqs.~(\ref{adin}), one makes the same assumptions as for 
Eqs.~(\ref{adout}) plus the assumption that $j_{\rm max}$ is small.

Comparison with the values listed in Tables \ref{tbl2} and \ref{tbl3}
shows that the adiabatic method agrees with the results of numerical 
integration usually to within 10\% or 20\%.  The overall agreement may 
be surprising considering that the first couple of throughfalls are 
not in the adiabatic regime, and that an exactly flat rotation curve 
is assumed to derive the predictions of the adiabatic approximation.  
Fig. \ref{rotcurv} shows that the rotation curve of the self-similar
infall model is only approximately flat, even for $\epsilon \simeq$ 0.3.

The adiabatic method is useful because it provides a check on the 
results of numerical integration.  It explains patterns in the 
numerical results that are otherwise mysterious.  In particular, 
it predicts that $v_n$ is $n$-independent and that $R_n$ and 
$a_n \propto {1 \over n}$ for large $n$; see Eqs.~(\ref{adout}) 
and (\ref{adin}).  These properties are descriptve of the 
values listed in Table \ref{tbl2} and \ref{tbl3}.  Finally, 
Eqs.~(\ref{adout}) and (\ref{adin}) enable quick estimates of 
all model properties for $\epsilon \neq 0.3$.

\section{Observational evidence}

In this section, we list a set of observations which are     
consistent with the caustic ring halo model, and thus lend  
it support.

\subsection{Flat rotation curves}

The model predicts the Galactic rotation curve at large radii $r$ 
all the way up to the turnaround radius $R$.  The rotation curves 
for various values of $\epsilon$ are shown in Fig. \ref{rotcurv} 
in dimensionless units.  They are nearly flat when $\epsilon \sim$
0.3, implying that the density $d$ is approximately proportional 
to ${1 \over r^2}$ at large $r$ all the way up to the turnaround 
radius $R$, whereabouts $d$ reaches the average value of the 
cosmological dark matter density.  In contrast, the density 
profiles predicted by computer simulations \cite{comsim} behave 
as $d(r) \propto {1 \over r^3}$ at large $r$, implying that the 
rotation velocity decreases as ${1 \over \sqrt{r}}$.

The rotation cuve of the Milky Way is consistent with 
being flat up to the largest radii, of order 20 kpc, 
where it has been measured.  More importantly, the rotation 
curves of spiral galaxies in general are flat at large radii 
up to the largest radii, of order 70 kpc, where they have been 
measured.  A collection of extended well measured rotation 
curves was published in refs. \cite{Bege,Sand}.  

Furthermore, studies of the dynamics of satellite galaxies 
\cite{Zari} provide evidence that the $d \propto r^{-2}$ 
behavior extends to $r \sim$ 200 kpc.  Weak lensing studies 
of the distortion of galaxy shapes by the gravitational fields 
of foreground galaxies \cite{Fish} provide evidence that the 
behavior extends to $r \sim$ 370 kpc.

\subsection{Effective core radii}

At small $r$, the Milky Way rotation curve is well accounted 
for by the baryonic matter in the Galactic bulge and disk.  So, 
the halo contribution to the rotation curve should be suppressed 
at small radii and should vanish at $r=0$ \cite{Galmod}.  The 
self-similar model generalized to include angular momentum of 
the dark matter particles \cite{STW} does have that property.
In ref. \cite{STW}, the `effective core radius' $r_{\rm c,eff}$ 
of a galactic halo was defined as the radius at which the halo 
contribution to the rotation velocity squared is half its value, 
$v_{\rm rot}^2$, at large $r$ where the halo dominates.  
$r_{\rm c,eff}$ is proportional to the average amount of 
dimensionless angular momentum $\bar{j}$.  The value of 
$r_{\rm c,eff}$ for the Milky Way, estimated by modeling 
the bulge and disk contributions to rotation curve, implies 
$\bar{j}_{\rm old} \sim 0.2$ \cite{STW}.

\subsection{Combined rotation curve}

Table \ref{tbl3} gives the model predictions for the caustic 
ring radii $a_n$.  The predictions may be approximated by 
\begin{equation}
a_n \simeq {40~{\rm kpc} \over n}~
\left({v_{\rm rot} \over 220~{\rm km/s}}\right)~
\left({j_{\rm max,old} \over 0.27}\right)
\label{crrf}
\end{equation}
for $\epsilon$ = 0.3.  For other values of $\epsilon$ in the 
range 0.2 to 0.35, the $a_n \propto {1 \over n}$ approximate 
law holds also, with the overall scale being $\epsilon$-dependent.  

Since the caustic rings are in or near the galactic plane, they 
cause bumps in the rotation curve at $r \simeq a_n$.  Galactic 
rotation curves have bumps for many reasons unrelated to caustics.  
However the bumps caused by caustics have a special pattern,
Eq.(\ref{crrf}).  One may hope to find evidence for this pattern 
in a statistical analysis of many galactic rotation curves.

In refs. \cite{Bege,Sand} 32 extended and well measured 
galactic rotation curves were selected under the criteria 
that each is an accurate tracer of the radial force law and 
that it extends far beyond the edge of the luminous disk.  
The data set of refs. \cite{Bege,Sand} was analyzed as follows
\cite{Kinn}. For each rotation curve, the radial variable was
rescaled according to  
\begin{equation}
r~\rightarrow~\tilde {r} = r
\left({220~{\rm km/s} \over v_{\rm rot}}\right)~~~\ ,
\label{resc}
\end{equation}  
where $v_{\rm rot}$ is the rotation velocity read off from that 
particular curve.   The data points for $\tilde{r} < 10$ kpc were 
deleted to remove the effect of the luminous disk.  The 32 rotation 
curves were then co-added to make a combined rotation curve.  

There are two peaks in the combined rotation curve, one 
near $40$ and one near $20\,{\rm kpc}$, with significance 
of $3.0\sigma$ and $2.6\sigma$, respectively.  No explanation 
has been given for the occurence of the two peaks other than that 
they are the effect of the $n=1$ and $n=2$ caustic rings of dark 
matter.  The result suggests not only the existence of caustic 
rings of dark matter distributed according to Eq. (\ref{crrf}), 
but also that the $j_{\rm max}$ distribution of the 32 galaxies 
in the data set is peaked near $j_{\rm max,old} = 0.27$ for 
$\epsilon = 0.3$.

That the $j_{\rm max}$ distribution is peaked near $j_{\rm max,old}$ = 
0.27 was not a prediction of the self-similar infall model but it is 
certainly an interesting outcome of the analysis.  It was mentioned 
in the previous subsection that the average magnitude $\bar{j}$ of 
the dimensionless angular momentum of the Milky Way had been determined 
to be $\bar{j}_{\rm old} \sim$ 0.2 from an estimate of the effective 
core radius of its halo.  For a given $j$ distribution on the turnaround 
sphere, $j_{\rm max}$  and $\bar{j}$ are related.  The simplest $j$ 
distribution is that of a rigidly rotating sphere 
[$j(\theta) = j_{\rm max} \sin \theta$], in which case $j_{\rm max} =
{4 \over \pi} \bar{j}$.  If we adopt this model, the estimate of 
the effective core radius of the Milky Way implies that its 
$j_{\rm max,old} \sim {4 \over \pi}$ 0.2 = 0.255 . This is close 
to the peak in the $j_{\rm max}$ distribution of the 32 external 
galaxies.  

Finally, we mention that the rotation curve of NGC3198, which may 
be the best measured in the set \cite{Bege,Sand}, by itself shows 
three faint bumps at radii consistent with Eq. (\ref{crrf}) 
and $j_{\rm max,old} = 0.28$ \cite{crdm}.

\subsection{Milky Way rotation curve}

The effect of a caustic ring in the plane of a galaxy upon its
rotation curve was analyzed in ref.\cite{sing}.  A caustic ring 
of radius $a$ and width $p$ produces a rise in the rotation curve 
which starts with an upward kink at $r_1 = a$ and ends with a 
downward kink at $r_2 = a + p$.  The two discontinuities are a 
direct consequence of the fact that the dark matter density 
associated with the caustic ring diverges at $r=a$ and $r=a+p$.

For technical reasons, the Milky Way rotation curve is
measured more precisely at $r < r_\odot$, where $r_\odot$
is our own galactocentric distance, than at $r > r_\odot$.
We assume the standard value $r_\odot = 8.5$ kpc.  The most 
detailed inner Galactic rotation curve, that we are aware of, 
was derived \cite{clem} from the Massachusetts-Stony Brook North
Galactic Plane CO survey \cite{MSB}.  It has a series of sharp 
rises between 3 and 8.5 kpc \cite{milk}.  Strikingly, each rise 
starts with an upward kink and ends with a downward kink, as 
expected for rises caused by caustic rings of dark matter.
Where each rise starts ($r=r_1$) and ends ($r=r_2$), the 
slope of the rotation curve changes abruptly, from one data 
point to the next.

Eq.~(\ref{crrf}) predicts ten caustic rings between 3 and 8.5 kpc, 
assuming $j_{\rm max,old}  \simeq 0.255$, as was inferred above 
from the effective core radius of the Milky Way halo.  Allowing
for ambiguities in identifying rises, the number of rises in the 
rotation curve \cite{clem} between 3 and 8.5 kpc is in fact ten 
plus or minus one.  The radii where the observed rises start 
($r_{1,n}$) and end ($r_{2,n})$ are listed in Table \ref{obscrp} 
under entries $n$ = 5,6 ... 14.  The self-similar infall model 
predictions for the caustic ring radii were fitted to the rises 
\cite{milk}.  For a given value of $\epsilon$, this is a one 
parameter fit, $j_{\rm max}$ being the only free parameter.  
For $\epsilon = 0.3$, the best fit occurs for $j_{\rm max,old} = 
0.263$.  This is consistent with the estimate $\bar{j}_{\rm old} 
\sim 0.2$ obtained from fitting the effective core radius of the 
Milky Way (see subsections IV.B and IV.C).  The root mean square 
relative deviation between the fitted caustic ring radii $a_n$ 
and the radii $r_{1n}$, where the rises in the rotation curve 
start with an upward kink, is 3.1\%.
  
As explained in Section III.A, $j_{\rm max,old} = 0.263$ in 
Einstein - de Sitter cosmology corresponds to $j_{\rm max} 
\simeq$ 0.179 in Concordance Cosmology.  We repeated the fit 
of the calculated $a_n$ to the observed $r_{1,n}$ in Concordance 
Cosmology and obtained $j_{\rm max}$ = 0.186, which is the value 
adopted here.  The slight shift in the fitted value of $j_{\rm max}$ 
compared to ref. \cite{milk} is due to the fact that the $a_n$ are 
only approximately proportional to $j_{\rm max}$, so that the 
overall increase in length scales by the factor 1.47 (see Section 
IIA) between the Einstein - de Sitter and Concordance cosmologies 
is only approximately compensated for by dividing $j_{\rm max}$ 
by 1.47.

As already mentioned, the rotation curve of the Milky Way is 
much  less well measured for $r > r_\odot$. It does however have 
a prominent rise between 12.7 and 13.7 kpc, where Eq.(\ref{crrf}) 
predicts the $n=3$ caustic ring to lie.  See, for example, the 
Milky Way rotation curve published in ref. \cite{Olli}.   

\subsection{Triangular feature in IRAS map}

The gravitational fields of caustic rings of dark matter may
leave imprints upon the distribution of ordinary matter.  Looking
tangentially to a caustic ring from a vantage point in the plane
of the ring, one may recognize the tricusp shape of the $D_{-4}$
catastrophe.  The IRAS map of the galactic disk in the direction
of galactic coordinates $(l,b) = (80^\circ, 0^\circ)$ shows a
triangular shape which is strikingly suggestive of the cross-section 
of a caustic ring \cite{milk}. The relevant IRAS maps are posted at 
http://www.phys.ufl.edu/$\sim$sikivie/triangle/~. The triangular 
shape is correctly oriented with respect to the galactic plane and 
the galactic center.  To an extraordinary degree of accuracy it is 
an {\it isosceles} triangle with axis of symmetry {\it parallel} to 
the galactic plane, as is expected for a caustic ring whose transverse 
dimensions are small compared to its radius.  Moreover its position is 
consistent with the position of the rise in the rotation curve, between
8.28 and 8.43 kpc, caused by the caustic ring of dark matter nearest 
to us.  

\subsection{The Monoceros Ring of stars}

The model predicts that the second caustic ring of dark matter 
($n=2$) lies in the Galactic plane at radius $a_2 \simeq$ 20 kpc.  
After this prediction was made \cite{crdm}, a ring of stars, named 
the ``Monoceros Ring", was discovered in the Galactic plane at 
$r \simeq$ 20 kpc \cite{Mono}.  It is shown in ref. \cite{Arav} 
that the Monoceros Ring of stars is the plausible outcome of the 
presence of the second caustic ring of dark matter.  In particular, 
it was shown that the adiabatic modification of disk star orbits 
by the gravitational field of the caustic ring  causes an order 
100\% enhancement of the density of disk stars at the location 
of the caustic ring.  It was also shown that viscous forces 
drive the gas in the neighborhood of the caustic ring towards 
the caustic ring radius $a_2$ which may therefore be a site of 
enhanced star formation.

\section{The flow densities and velocities on Earth}

Our study of the phase space structure of the Milky Way halo 
is motivated in large part by the ongoing searches for dark 
matter on Earth, using axion \cite{axion} and WIMP \cite{WIMP}
detectors.  The signal in such detectors depends on the velocity 
distribution of dark matter in the solar neighborhood.  The caustic 
ring halo model predicts that most of the local dark matter is in 
discrete flows, and predicts the velocity vectors and densities 
of the flows.  Table \ref{localflows} lists the properties of the 
first forty flows at the Earth's location in the Galaxy.  Earlier 
versions of this table appeared in refs. \cite{IDM98} and \cite{annual}.  
The purpose of this section is to describe how the estimates of Table
\ref{localflows} were obtained and to comment on the uncertainties.

As described earlier, the flows come in pairs labeled $(n,\pm)$.
The flow velocities are stated in a reference frame which is 
attached to the Galaxy but which is not rotating relative to 
the faraway universe.  The flow speeds (column 2) follow from 
the kinetic energy acquired by the particles while they fall 
in the growing gravitational potential well of the Galaxy.  
We calculate the speeds at the location of the Sun, at 
$r_\odot \equiv$ 8.5 kpc from the Galactic center, in 
the spherically symmetric self-similar model.  Thus, 
\begin{equation}
v_{n\pm}(r_\odot) = {R \over t}~\xi_{n\pm}~ 
{d \Lambda \over d\xi} (\xi_{n\pm})
\label{Espeeds}
\end{equation}
where $R$ = 2.4 Mpc, $t$ = 13.7 Gyr, and the $\xi_{n\pm}$ 
are the solutions of:
\begin{equation}
r_\odot = R \Lambda(\xi)~~~\ .
\label{xisol}
\end{equation}
There are two solutions for each $n$ but the corresponding speeds 
are nearly equal.  The difference between $v_{n+}(r_\odot)$ and 
$v_{n-}(r_\odot)$ is ignored in Table \ref{localflows}. 

The velocity components are listed in columns 3 through 5 of 
Table \ref{localflows}, and the densities in columns 6 and 7.  
The previous sections provide two ways to estimate the flow 
velocities and densities.  For those flows whose inner caustic 
rings are close to us, we have the description of Section 
IIIC.  For those flows whose inner caustic rings are far 
from us, we have the description of Section IIID, and 
specifically the approximate equations (\ref{exap}).  
In either case, we assume that the Sun is in the 
symmetry plane ($z=0$) of the model.  This leads 
to some simplifications.

For $z=0$ Eqs.~(\ref{crfl}), which describe the flow near a 
caustic ring, have two pairs of solutions, corresponding to 
$\alpha = 0$ and $\eta = 0$.  For $\alpha = 0$ we have:
\begin{eqnarray}
r = \rho &=& a + {1 \over 2} u (\eta - \eta_0)^2\nonumber\\
v_z &=& 0\nonumber\\
v_\rho = - {\partial \rho \over \partial \eta} &=&
- u (\eta - \eta_0) = \mp v \sqrt{2({r \over a} -1)}\nonumber\\
d_\pm &=& {1 \over r} {dM \over d\Omega d t}~
{1 \over 2 b \sqrt{r - a} |\sqrt{r - a} \pm \sqrt{p}|}~~~\ .
\label{alpha0}
\end{eqnarray}  
The choice of sign in the equation for $v_\rho$ is independent 
of the choice of sign in the equation for $d_\pm$.  Eqs.~(\ref{alpha0}) 
describe the in and out flows which exist in the plane of the ring for 
$r>a$.  For $\eta = 0$ we have
\begin{eqnarray}
r = \rho &=& a + p - {1 \over 2} s \alpha^2\nonumber\\
v_z = - {\partial z \over \partial \eta} &=& - b \alpha
= \mp b \sqrt{{2 \over s} (a + p - r)}\nonumber\\
v_\rho = - {\partial \rho \over \partial \eta} &=& + u \eta_0
= \pm v \sqrt{2p \over a}\nonumber\\
d_\pm &=& {1 \over r} {dM \over d\Omega d t}~
{\cos \sqrt{{2 \over s} (a + p - r)} \over 2 b (a + p - r)}~~~\ .
\label{eta0}
\end{eqnarray}
The choice of sign in the equation for $v_z$ is independent of 
the choice of sign in the equation for $v_\rho$.  Eqs. (\ref{eta0}) 
describe the up and down flows which exist in the plane of the ring 
for $r < a + p$.

For $z=0$ Eqs. (\ref{exap}) become
\begin{eqnarray}
\vec{v} &=& \hat{\varphi}~v~{r \over a} 
\pm \hat{z}~v~\sqrt{1 - \left({r \over a}\right)^2}~~~~~~~{\rm for}
~~r<a\nonumber\\ &=& \hat{\varphi}~v~{a \over r}
\pm \hat{\rho}~v~\sqrt{1 - \left({a \over r}\right)^2}~~~~~~~~~
{\rm for}~~r>a\nonumber\\
d_\pm &=& {1 \over v} {dM \over d\Omega d t} {1 \over |r^2 - a^2|}~~~\ .
\label{flz0}
\end{eqnarray} 
The (only) justification for Eqs.~(\ref{exap}), and hence for 
Eqs.~(\ref{flz0}), is that they provide a smooth interpolation 
of the flows between the inner and outer caustics in the limit 
where the transverse dimensions, $p$ and $q$, of the inner caustic 
are neglected.  For flows of $n$ near 5, whose caustic rings 
are near us, Eqs.~(\ref{alpha0}) and (\ref{eta0}) provide a more 
accurate description but they involve the parameters $b$ and $s$ 
about which we have little information other than that $b$ is of 
order $v$, and $s$ of order $a$.  Eqs.~(\ref{exap}) and
(\ref{flz0}) merely assume $b=v$ and $s=a$.  One may verify 
that, for $b=v$ and $s=a$, Eqs.~(\ref{flz0}) are consistent with 
Eqs.~(\ref{alpha0}) and (\ref{eta0}) when $r$ is near $a$ but $p=0$.

The velocity components listed in columns 3 through 5 of 
Table \ref{localflows} were obtained using Eqs.~(\ref{flz0}).
For those values of $n$ (i.e. $n$ = 3, 5 through 14) for which 
there is evidence for caustic rings of dark matter in the form 
of sharp rises in the Galactic rotation curve, we used the caustic 
ring radii $r_{1,n}$ inferred from the positions of the rises.  
The $r_{1,n}$ are listed in Table \ref{obscrp}.  For the other 
values of $n$ (i.e. $n$ = 1, 2, 4, 15 through 20) we used the 
predicted values listed in Table \ref{tbl3}.  

Eqs.~(\ref{flz0}) predict that $v_z^{n\pm}(\vec{r}_\odot)$ = 0 for 
$n \geq 5$ and $v_\rho^{n\pm}(\vec{r}_\odot)$ = 0 for $n \leq 4$.  
That $v_z^{n\pm}(\vec{r}_\odot)$ = 0 for $n \geq 5$ follows from 
the symmetry of the model.  We have set the corresponding 
entries equal to zero in Table \ref{localflows}.  On the 
other hand, that $v_\rho^{n\pm}(\vec{r_\odot})$ = 0 for 
$n \leq 4$ is a consequence of the approximations that led
to Eqs.~(\ref{exap}), in particular the approximation $p=0$.  
We may compare the zero value for $v_\rho$ in Eqs.~(\ref{flz0}) 
for $r<a$ with the expression for $v_\rho$ in Eqs. (\ref{eta0}) 
in terms of quantities $v$ and $a$, which we believe we know well, 
and $p$ on which there is observational information.   The widths 
$p_n$ of those caustic rings for which there is evidence in the 
form of sharp rises in the Galactic rotation curve may be obtained 
from Table IV by setting $p_n = r_{2,n} - r_{1,n}$. For this set, the 
ratio ${p \over a}$ varies from 0.015 to 0.10, with an average of 0.05.  
Eqs.~(\ref{eta0}) imply therefore that, at least for $r$ near $a$, the 
up and down flows for $r < a+p$ have $|v_\rho| \sim 0.3~v$, whereas
Eqs.~(\ref{flz0}) would predict a zero value.  As a reminder of 
these uncertainties, the entries in Table \ref{obscrp} for 
$v^{n\pm}_\rho(\vec{r}_\odot)$ with $n < 5$ have been left 
blank.

The flow densities listed in the last two columns of Table 
\ref{localflows} were calculated using Eqs.~(\ref{flz0}) for 
$n \leq 4$ and $n \geq 10$, and using Eqs.~(\ref{alpha0}) 
with $b = v$ for $5 \leq n \leq 9$.  Again we used the 
observed values $r_{1,n}$ of the caustic ring radii for 
$n$ = 3, 5 through 14 and the predicted $a_n$ for $n$ = 1,2
4 and 15 through 20.  The values of ${d M \over d \Omega d t}$ 
were taken from the last column of Table \ref{tbl3}.  Since the 
ratio ${v \over b}$ was set equal to one, whereas it is only known 
to be of order one, {\it all density estimates are uncertain by at 
least a factor two.} 

\section{Summary}

In this paper our main goal has been to provide a complete and 
self-contained description of the caustic ring model of the Milky 
Way halo, so that anyone wishing to compare the model with observations 
may readily obtain its predictions.  Aside from the assumption of cold
collisionless dark matter, the defining properties of the model are 
axial and reflection symmetry, self-similarity and net overall rotation.  
The model is a proposal for the full phase space distribution of the 
Milky Way halo, which is described as a set of discrete flows with 
stated densities and velocities everywhere.

In Section II, we discussed the general conditions under
which the time evolution of a self-gravitating cold 
collisionless fluid is self-similar.  We found that
self-similarity does not require symmetry.  We discussed
three cases: no symmetry, spherical symmetry, and axial 
symmetry.  We derived the equations of motion for the 
reduced variables in each instance, and the initial 
conditions consistent with self-similarity.

The model breaks spherical symmetry because the dark matter 
particles carry angular momentum.  The angular momentum 
distribution determines the properties of the inner parts 
of the halo, and in particular of the inner caustics.  
However, on large scales, the model is nearly spherically 
symmetric.  The spherically symmetric model was originally 
described by Fillmore and Goldreich \cite{FG}, and by 
Bertschinger \cite{B}.  It depends on one dimensionless 
parameter $\epsilon$ and two dimensionful parameters.  The 
latter merely set the scale of the halo's extent in phase 
space.  They are determined in our model by requiring the 
Galactic rotation velocity at the location of the Sun to 
be 220 km/s and the age of the Galaxy to be 13.7 Gyr.  The 
parameter $\epsilon$ is related to the slope of the power 
spectrum of density perturbations on galactic scales.  This 
implies $\epsilon \sim 0.3$.  The properties of the model 
fitted to observation do not depend sharply on $\epsilon$.  
We set $\epsilon$ = 0.3 for definiteness. 

The properties of the outer caustic spheres, the flow speeds
and the mass infall rates were calculated in the spherically 
symmetric model.  The properties of the first twenty outer 
caustics are listed in Table \ref{tbl2}.  The flow speeds at 
the position of the Sun are listed in Table \ref{localflows}.  
The flow speeds at the location of each flow's inner caustic 
are listed in Table \ref{tbl3}.  The mass infall rates are 
given by Eq.~(\ref{infre}).  Their values at the inner caustic 
radii are given in Table \ref{tbl3}.  

The structure of the inner caustics depends on the angular 
momentum distribution of the infalling dark matter particles.
The additional parameters introduced in our axially symmetric 
and self-similar halo model, relative to the spherically symmetric 
model, are the components $\nu_r(\alpha)$, $\nu_\theta(\alpha)$ 
and $\nu_\varphi(\alpha)$ of the rescaled initial velocity distribution 
on the turnaround sphere, Eq.~(\ref{axsinv}).  The radial component
$\nu_r(\alpha)$ only plays a minimal role \cite{inner} because it 
does not contribute to angular momentum.  We assume that the angular 
momentum distribution is such that there is net overall rotation, i.e. 
$\nu_\varphi(\alpha)$ dominates over $\nu_\theta(\alpha)$.  We make 
this assumption because in that case the inner caustics are tricusp 
rings.  There is evidence for such rings in the Milky Way and other 
isolated spiral galaxies.  The evidence is summarized in Section IV.  
The radii of the caustic rings depend on a single parameter 
$j_{\rm max} \equiv \nu_\varphi(0)$.  The evidence is consistent 
with $j_{\rm max}$ = 0.186, the value adopted here.  The caustic 
ring radii are listed in Table \ref{tbl3}.

The flow velocities and densities in the neighborhood of a caustic 
ring are described in Section IIIC in terms of 6 parameters which 
may be chosen to be: the caustic ring radius $a_n$, the flow speed 
$v_n$ at the caustic, the mass infall rate ${d M \over d\Omega d\eta}|_n$ 
there, the time scale $\eta_{0,n}$ over which the particles traverse
the caustic, a parameter $b_n$ of order $v_n$, and a parameter $s_n$
of order $a_n$.  The model values of the first three parameters are 
given in Table \ref{tbl3} for the first twenty inner caustics.  The 
last three parameters ($\eta_{0,n}$, $b_n$ and $s_n$) are related 
to relatively subtle details of the initial velocity distribution, 
as described in subsection III.C.4.  They determine the transverse
sizes $p_n$ and $q_n$ of the caustic rings (see Eq.~(\ref{pq})), and 
also enter the formula for the density Eq.~(\ref{denax}). 
 
The model properties are obtained by solving a couple of equations 
(e.g. Eq.~(\ref{eom3})) on a computer.  However, quick estimates 
can be gotten by using the method of adiabatic invariants.  The 
main results of that method are stated in Eqs.~(\ref{adout}) and 
(\ref{adin}).  The method of adiabatic invariants usually agrees 
with the results of numerical integration to within 10\% or 20\%.  
It allows estimates of the model properties for $\epsilon \neq$ 0.3.

Table \ref{localflows} lists the predicted densities and velocity 
vectors of the first forty flows on Earth.  Knowledge of the local 
dark matter velocity distribution is essential when interpreting a 
signal in a dark matter detector on Earth.  A striking property 
of the model is the existence on Earth of a "Big Flow", entry 5+ 
in Table \ref{localflows}, whose existence should become readily 
apparent if a signal is found.  The model makes predictions for 
all approaches to the dark matter problem, including direct searches 
for axions \cite{annual,axion} and WIMPs \cite{WIMPphen,annual},
strong and weak gravitational lensing \cite{gravlens}, WIMP 
annihilation in the Galactic halo \cite{annih} and the cosmic 
rays produced thereby \cite{Salati}.

%\acknowledgments (R3)
\begin{acknowledgments}
We are grateful to Igor Tkachev for having made available 
to us his numerical codes for solving the equations of the 
self-similar model.  This work was supported in part by the
U.S. Department of Energy under grant DE-FG02-97ER41209.  
P.S. gratefully acknowledges the hospitality of the Aspen 
Center for Physics while working on this project.
\end{acknowledgments}

%%%%%%%%%%%%%%%%%%%%%%%%%%%%%%%%%%%%%%%%%%%%%%%

\newpage

\begin{table}
\caption{The dimensionless rotation velocity squared at the Sun's 
position $\nu^2(\epsilon, {r_\odot \over R})$, the turnaround radius 
$R$ and the total mass $M$ for different values of $\epsilon$.}
\begin{tabular}{|c|c|c|c|}
\hline
 $~~~~\epsilon$~~~~&~~~~$\nu^2(\epsilon,{r_\odot \over R})$~~~~& 
~~~~$R$~(Mpc)~~~~&~~~~$M~(M_\odot)$~~~~\\
\hline
 0.1   & 0.21  & 6.0 & $3.1\times10^{14}$  \\
 0.15  & 0.51  & 3.9 & $8.4\times10^{13}$  \\
 0.2   & 0.81  & 3.1 & $4.3\times10^{13}$  \\
 0.25  & 1.08  & 2.7 & $2.8\times10^{13}$  \\
 0.3   & 1.35  & 2.4 & $2.0\times10^{13}$  \\
 0.35  & 1.62  & 2.2 & $1.5\times10^{13}$  \\
 0.4   & 1.91  & 2.0 & $1.2\times10^{13}$  \\
 0.45  & 2.15  & 1.9 & $9.8\times10^{12}$  \\
\hline
\end{tabular}
\label{tbl1}
\end{table}

\begin{table}
\caption{The radii $R_n$ and fold coefficients $A_n$
of the first 20 outer caustics.}
\begin{tabular}{|c|c|c|}
\hline
~~~$n$~~~&~~~~$R_n$ (kpc)~~~~&
~~~~$A_n~(M_\odot / {\rm pc}^{5 \over 2})$~~~~\\
\hline
 1   & 436  & $7.7\times10^{-4}$  \\
 2   & 260  & $8.8\times10^{-4}$  \\
 3   & 187  & $9.6\times10^{-4}$  \\
 4   & 147  & $1.02\times10^{-3}$  \\
 5   & 121  & $1.07\times10^{-3}$  \\
 6   & 103  & $1.12\times10^{-3}$  \\
 7   & 89  & $1.16\times10^{-3}$  \\
 8   & 79  & $1.20\times10^{-3}$  \\
 9   & 71  & $1.24\times10^{-3}$  \\
10   & 64  & $1.27\times10^{-3}$  \\
11   & 59  & $1.29\times10^{-3}$  \\
12   & 54  & $1.32\times10^{-3}$  \\
13   & 50  & $1.35\times10^{-3}$  \\
14   & 47  & $1.37\times10^{-3}$  \\
15   & 44  & $1.40\times10^{-3}$  \\
16   & 41  & $1.42\times10^{-3}$  \\
17   & 39  & $1.44\times10^{-3}$  \\
18   & 37  & $1.46\times10^{-3}$  \\
19   & 35  & $1.48\times10^{-3}$  \\
20   & 34  & $1.50\times10^{-3}$  \\
\hline
\end{tabular}
\label{tbl2}
\end{table}

\begin{table}
\caption{The radii $a_n$, flow speeds $v_n$ and infall
rates ${d M \over d \Omega d \eta}|_n$ for the first 20 
inner caustics.}
\begin{tabular}{|c|c|c|c|}
\hline
~~~$n$~~~ &~~~~~$a_n$~~~~~&~~~~~$v_n$~~~~~& 
${d M \over d \Omega d \eta}|_n$~\\
~~~&~(kpc)~&~(km/s)~&$({M_\odot \over {\rm sterad}\cdot{\rm yr}})$\\
\hline
 1   & 40.1  & 517  & 53 \\
 2   & 20.1  & 523  & 23 \\
 3   & 13.6  & 523  & 14 \\
 4   & 10.4 & 523  & 10 \\
 5   & 8.4  & 522  & 7.8 \\
 6   & 7.0  & 521  & 6.3 \\
 7   & 6.1  & 521  & 5.3 \\
 8   & 5.3  & 520  & 4.5 \\
 9   & 4.8  & 517  & 3.9 \\
10   & 4.3  & 515  & 3.4 \\
11   & 4.0  & 512  & 3.1 \\
12   & 3.7  & 510  & 2.8 \\
13   & 3.4  & 507  & 2.5 \\
14   & 3.2  & 505  & 2.3 \\
15   & 3.0  & 503  & 2.1 \\
16   & 2.8  & 501  & 2.0 \\
17   & 2.7  & 499  & 1.8 \\
18   & 2.5  & 497  & 1.7 \\
19   & 2.4  & 496  & 1.6 \\
20   & 2.3  & 494  & 1.5 \\
\hline
\end{tabular}
\label{tbl3}
\end{table}

\begin{table}
\caption{Galactocentric radii where observed rises in the 
Milky Way rotation curves start ($r_{1,n}$) and end ($r_{2,n}$).}
\begin{center}
\footnotesize
\begin{tabular}{|c|c|c|}
\hline
$~~n~~$&$~~~r_{1,n}$~(kpc)~~~&$~~~r_{2,n}$~(kpc)~~~\\
\hline
3 & 12.7 & 13.7 \\
5 & 8.28 & 8.43 \\
6 & 7.30 & 7.42 \\
7 & 6.24 & 6.84 \\
8 & 5.78 & 6.01 \\
9 & 4.91 & 5.32 \\
10 & 4.18 & 4.43 \\
11 & 3.89 & 4.08 \\
12 & 3.58 & 3.75 \\
13 & 3.38 & 3.49 \\
14 & 3.16 & 3.25 \\
\hline
\end{tabular}
\end{center}
\label{obscrp}
\end{table}

\begin{table}
\caption{Velocity vectors $\vec{v}^{n\pm}(\vec{r}_\odot)$ 
and densities $d_n^\pm(\vec{r}_\odot)$ of the first 40 flows 
at our location in the Milky Way, in galactic coordinates.  
$\hat{\varphi}$ is in the direction of galactic rotation,
$\hat{z}$ is perpendicular to the galactic plane, and 
$\hat{\rho}$ is in the radial direction.  The two flows 
with velocity vector $\vec{v}^{n\pm}$ have density 
$d_n^\pm$ or $d_n^\mp$, for each $n$ independently.
The $v^{n\pm}_\rho(\vec{r}_\odot)$ entries for 
$n \leq$ 4 are not predicted but are expected to 
be relatively small; see text.}
\begin{center}
\footnotesize
\begin{tabular}{|r|c|c|c|c|c|c|}
\hline
$n$&$v^{n\pm}(\vec{r}_\odot)~~~$&$v^{n\pm}_\varphi(\vec{r}_\odot)$~~~&
$v^{n\pm}_z(\vec{r}_\odot)~~~$&$v^{n\pm}_\rho(\vec{r}_\odot)~~~$&
$d_n^{+}(\vec{r}_\odot)~$&$d_n^{-}(\vec{r}_\odot)~$\\
&(km/s)~~~&(km/s)~~~&(km/s)~~~&(km/s)~~~&
($10^{-26}$gr/cm$^3$)~~&($10^{-26}$gr/cm$^3$)~~\\
\hline
1 & 650 & 140 & $\pm$ 635 & / & 0.3 & 0.3 \\
2 & 600 & 250 & $\pm$ 540 & / & 0.8 & 0.8 \\
3 & 565 & 380 & $\pm$ 420 & / & 1.9 & 1.9 \\
4 & 540 & 440 & $\pm$ 310 & / & 3.4 & 3.4 \\
5 & 520 & 505 & 0 & $\pm$ 120 & 150 & 15 \\
6 & 500 & 430 & 0 & $\pm$ 260 & 6.0 &  3.1 \\
7 & 490 & 360 & 0 & $\pm$ 330 & 3.9 &  1.2 \\
8 & 475 & 325 & 0 & $\pm$ 350 & 1.9 &  1.0 \\
9 & 460 & 265 & 0 & $\pm$ 375 & 1.4 &  0.7 \\
10 & 450 & 220 & 0 & $\pm$ 390 & 0.9 & 0.9 \\
11 & 440 & 200 & 0 & $\pm$ 390 & 0.8 & 0.8 \\
12 & 430 & 180 & 0 & $\pm$ 390 & 0.7 & 0.7 \\
13 & 420 & 170 & 0 & $\pm$ 390 & 0.6 & 0.6 \\
14 & 415 & 155 & 0 & $\pm$ 385 & 0.6 & 0.6 \\
15 & 405 & 140 & 0 & $\pm$ 380 & 0.5 & 0.5 \\
16 & 400 & 130 & 0 & $\pm$ 375 & 0.5 & 0.5 \\
17 & 390 & 120 & 0 & $\pm$ 370 & 0.5 & 0.5 \\
18 & 380 & 110 & 0 & $\pm$ 365 & 0.4 & 0.4 \\
19 & 375 & 100 & 0 & $\pm$ 360 & 0.4 & 0.4 \\
20 & 370 &  95 & 0 & $\pm$ 355 & 0.4 & 0.4 \\
\hline
\end{tabular}
\end{center}
\label{localflows}
\end{table}

%%%%%%%%%%%%%%%%%%%%%%%%%%%%%%%%%%%%%%%%%%%%%%

\newpage

%\begin{references} (R3)

%%%%%%%%%%%%%%%%%%%%%%%%%%%%%%%%%%%%%%%%%%%%%%%

\newpage

\begin{figure}
\begin{center}
\resizebox{15cm}{10cm}{\includegraphics{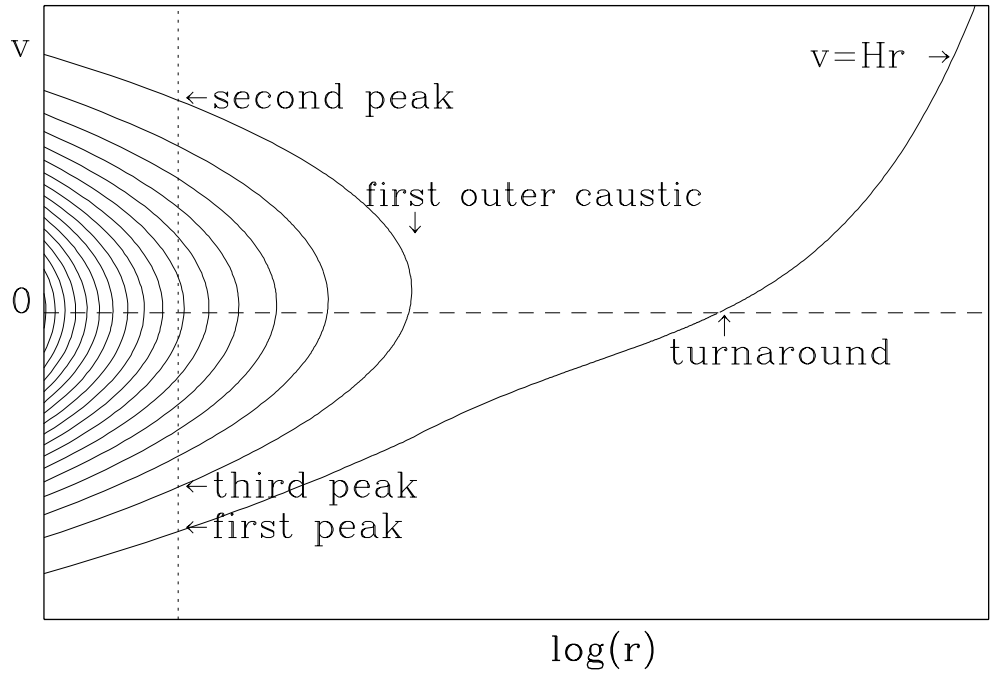}}
\caption{Qualitative description of the phase space distribution 
of dark matter particles in a halo at a fixed moment of time. The 
solid lines represent occupied phase space cells.  The vertical 
dotted line gives the observer position.  Each intersection of 
the solid and dotted lines corresponds to a dark matter flow at 
the observer position.} 
\end{center}
\label{phase}
\end{figure}

\begin{figure}
\resizebox{15cm}{8.5cm}{\includegraphics{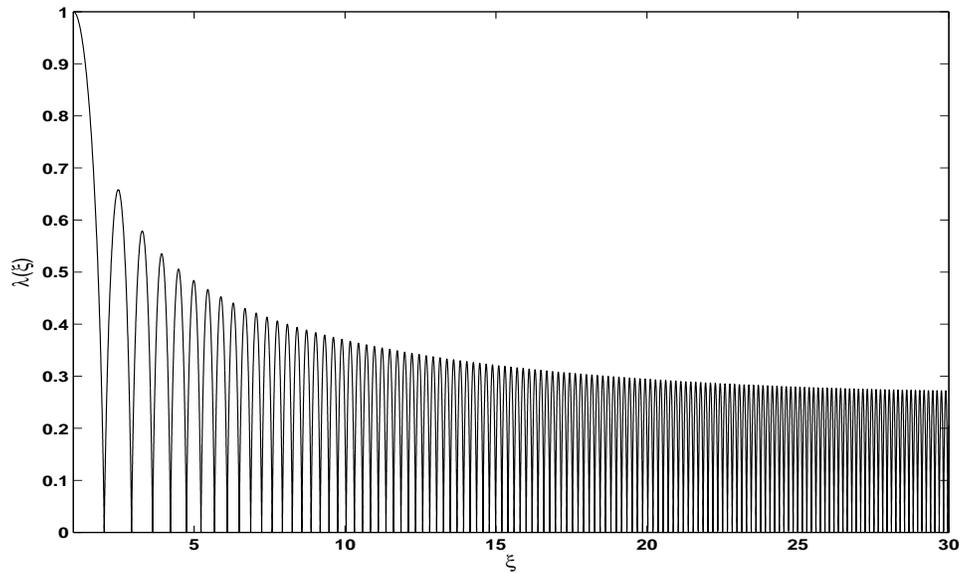}}
\caption{The function $\lambda (\xi)$ in the spherically 
symmetric self-similar infall model with $\epsilon =0.3$.}
\label{lambda}
\end{figure}

\begin{figure}
\resizebox{15cm}{10cm}{\includegraphics{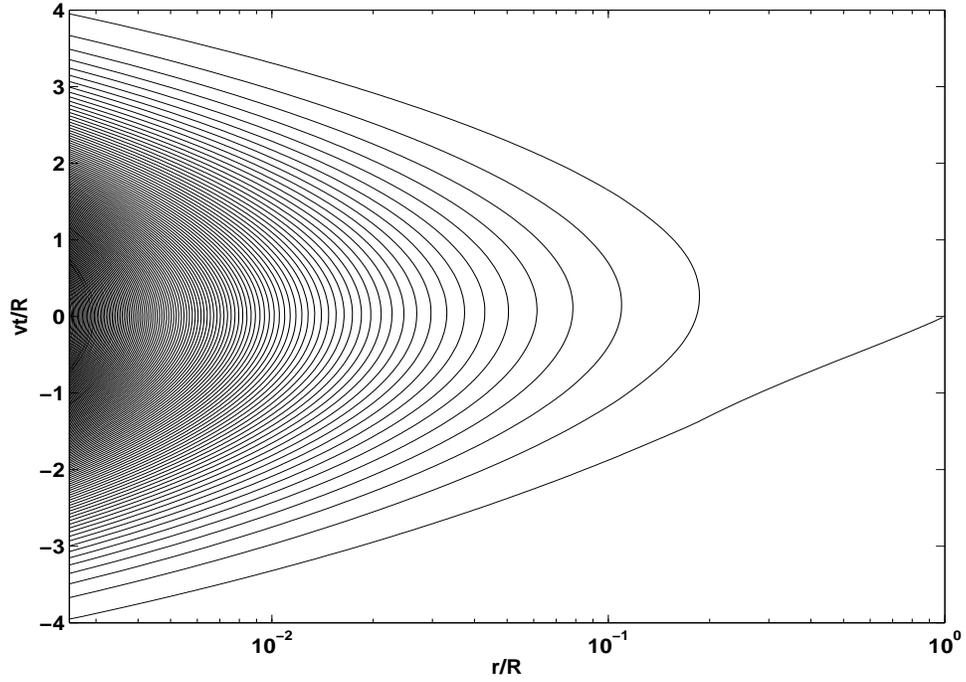}}
\caption{The phase space distribution of halo dark matter particles 
at a fixed moment of time in the spherically symmetric self-similar
infall model with $\epsilon =0.3$. The solid lines represent occupied
phase space cells.}
\label{selfsim}
\end{figure}

\begin{figure}
\resizebox{15cm}{10cm}{\includegraphics{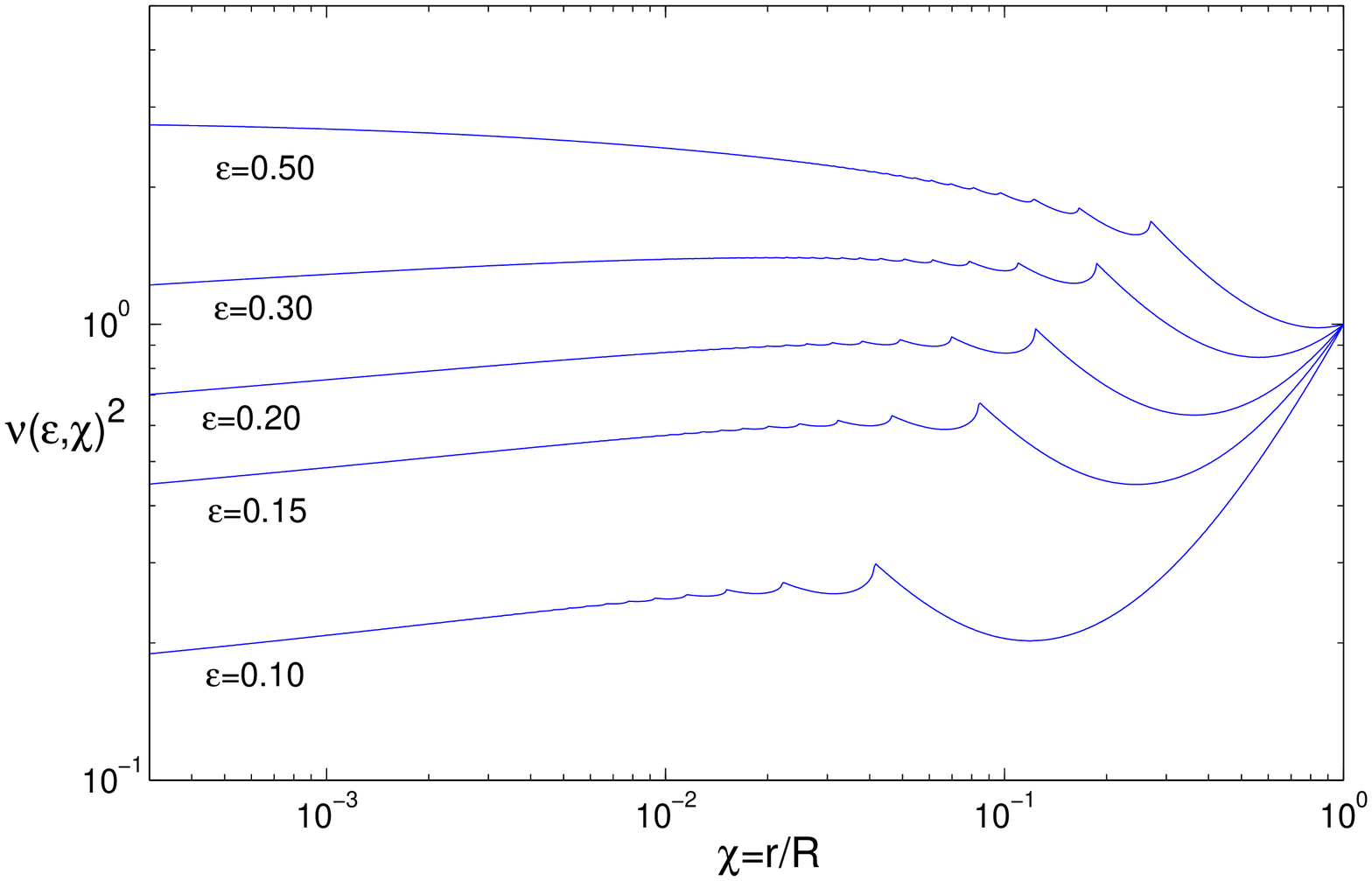}}
\caption{Rotational velocity squared curves for different values of
$\epsilon$, and $j=0$.}
\label{rotcurv}
\end{figure}

\begin{figure}
\centering
\resizebox{9cm}{9cm}{\includegraphics{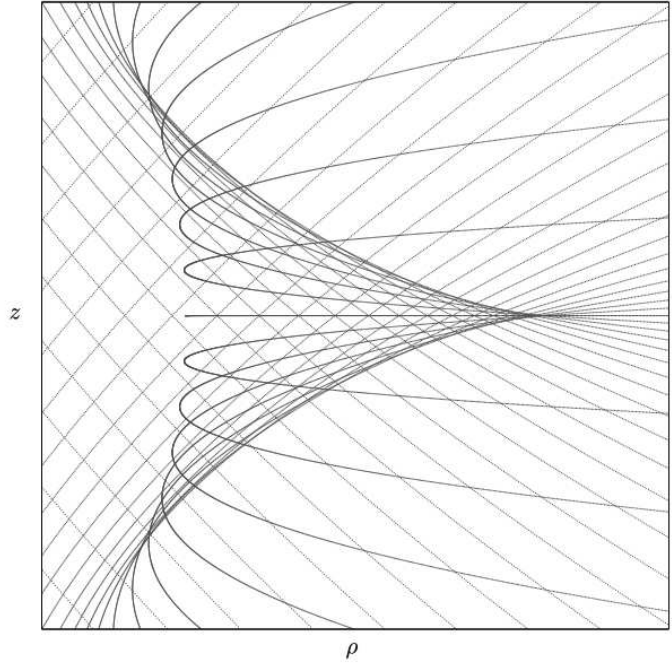}}
\caption{Dark matter trajectories forming a caustic ring of dark matter, 
in $\rho$-$z$ cross-section.}
\label{flow}
\end{figure}

\begin{figure}
\resizebox{9cm}{9cm}{\includegraphics{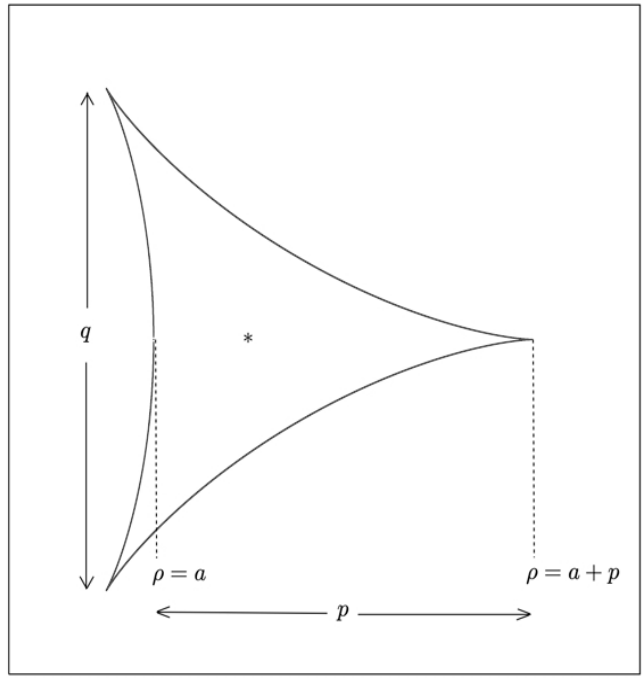}}
\caption{The envelope of the trajectories that are shown in
Fig.~\ref{flow}.  The density diverges at the envelope in 
the limit of zero velocity dispersion. We refer to the shape 
shown as the ``tricusp".  What is meant by the radius $a$ and 
the transverse sizes $p$ and $q$ of a caustic ring is indicated.
The tricusp has a discrete symmetry involving a rescaling and 
rotation by $120^\circ$ about the central point, indicated by 
a star.}
\label{dimen}
\end{figure}

\end{document}